\documentclass[acmsmall,screen]{acmart} 

\makeatletter
\def\@mkauthorsaddresses{%
  \bgroup
  \def\streetaddress##1{\ClassWarning{\@classname}{ACM no longer collects
  authors' postal addresses. I am ignoring your street address}\unskip\ignorespaces}%
  \def\postcode##1{\ClassWarning{\@classname}{ACM no longer collects
  authors' postal addresses. I am ignoring your postal code}\unskip\ignorespaces}%
  \def\position##1{\unskip\ignorespaces}%
  \gdef\@ACM@institution@separator{, }%
  \def\institution##1{\unskip\@ACM@institution@separator ##1%
    \gdef\@ACM@institution@separator{ and }}%
  \def\city##1{\unskip, ##1}%
  \def\state##1{\unskip, ##1}%
  \renewcommand\department[2][0]{\unskip\@addpunct, ##2}%
  \def\country##1{\unskip, ##1}%
  \def\and{\unskip; \gdef\@ACM@institution@separator{, }}%
  \def\@author##1{##1}%
  \newif\if@firstemail \@firstemailtrue
  \def\email##1##2{%
    \if@firstemail
      Emails: \nolinkurl{##2}\@firstemailfalse
    \else
      , \nolinkurl{##2}%
    \fi
  }%
  \addresses
  \egroup
}
\makeatother

\setcopyright{none}
\renewcommand\footnotetextcopyrightpermission[1]{}

\AtBeginDocument{%
  }

\usepackage{xcolor}
\usepackage{pifont}
\usepackage{adjustbox}
\usepackage{booktabs}
\usepackage{multirow}
\usepackage{colortbl}
\usepackage{enumitem}

\usepackage{tikz}

\usepackage{forest}
\usetikzlibrary{shadows}
\usepackage{graphicx}
\usepackage{amsmath, amssymb}
\usepackage{cleveref}
\usepackage{float}
\usepackage{subcaption}

\newcommand{\cmark}{\textcolor{green!60!black}{\checkmark}}        
\newcommand{\hmark}{\textcolor{orange!85!black}{\small$\triangle$}}
\newcommand{\xmark}{\textcolor{red!75!black}{\ding{55}}}           

\begin{document}

\title{Copyright Protection for Large Language Models: A Survey of Methods, Challenges, and Trends}

\email{xuzhenhua0326@zju.edu.cn, yuexubin@zju.edu.cn, breynald@zju.edu.cn, kdz@zju.edu.cn, wpxing@zju.edu.cn, mhan@zju.edu.cn, xixiangzhao77@gmail.com, linchangting@gmail.com, qichen.liu@alumni.wfu.edu, e1349386@u.nus.edu, jz97@iu.edu}

\thanks{\textsuperscript{$\ast$}Equal contribution.}
\thanks{\textsuperscript{$\dagger$}Corresponding author.}

\renewcommand{\shortauthors}{Zhenhua Xu, Xubin Yue, Zhebo Wang, Haobo Zhang, Xixiang Zhao, Qichen Liu et al.}

\begin{abstract}

\vspace{-15pt}
\begin{center}
{\small
Zhenhua Xu\textsuperscript{$1\ast$},
Xubin Yue\textsuperscript{$1\ast$},
Zhebo Wang\textsuperscript{$1\ast$},
Haobo Zhang\textsuperscript{$2\ast$},
Xixiang Zhao\textsuperscript{$2\ast$},
Qichen Liu\textsuperscript{$2\ast$} \\
Jingxuan Zhang\textsuperscript{$2$}, 
Wenjun Zeng\textsuperscript{$2$},
Wenpeng Xing\textsuperscript{$1,2$},
Dezhang Kong\textsuperscript{$1,2$},
Changting Lin\textsuperscript{$1$},
Meng Han\textsuperscript{$1\dagger$},
}\\[6pt]
\textsuperscript{$1$}Zhejiang University \quad
\textsuperscript{$2$}Binjiang Institute of Zhejiang University
\end{center}

\noindent \textbf{Abstract:} Copyright protection for large language models is of critical importance, given their substantial development costs, proprietary value, and potential for misuse. Existing surveys have predominantly focused on techniques for tracing LLM-generated content—namely, text watermarking—while a systematic exploration of methods for protecting the models themselves (i.e., model watermarking and model fingerprinting) remains absent. Moreover, the relationships and distinctions among text watermarking, model watermarking, and model fingerprinting have not been comprehensively clarified. This work presents a comprehensive survey of the current state of LLM copyright protection technologies, with a focus on model fingerprinting, covering the following aspects: (1) clarifying the conceptual connection from text watermarking to model watermarking and fingerprinting, and adopting a unified terminology that incorporates model watermarking into the broader fingerprinting framework; (2) providing an overview and comparison of diverse text watermarking techniques, highlighting cases where such methods can function as model fingerprinting; (3) systematically categorizing and comparing existing model fingerprinting approaches for LLM copyright protection; (4) presenting, for the first time, techniques for fingerprint transfer and fingerprint removal; (5) summarizing evaluation metrics for model fingerprints, including effectiveness, harmlessness, robustness, stealthiness, and reliability; and (6) discussing open challenges and future research directions.  
This survey aims to offer researchers a thorough understanding of both text watermarking and model fingerprinting technologies in the era of LLMs, thereby fostering further advances in protecting their intellectual property. We will continue to maintain and update a curated list of related papers and resources at \href{https://github.com/Xuzhenhua55/awesome-llm-copyright-protection}{\texttt{https://github.com/Xuzhenhua55/awesome-llm-copyright-protection}}.
\end{abstract}


\keywords{large language models,copyright protection, text watermarking,model fingerprinting}


\maketitle

\section{Introduction}
\label{introduction}

Over the past decade, deep learning has undergone remarkable evolution, progressing from early convolutional and recurrent neural network architectures~\cite{goodfellow2016deep, lecun2015deep, rawat2017deep, medsker1999recurrent} to today's transformative large language models (LLMs)~\cite{lalai2024intentions}. These models, underpinned by the advances in transformer architectures and unprecedented access to large-scale training data, have demonstrated an extraordinary capacity for a wide range of natural language processing (NLP) tasks. Modern LLMs—such as GPT-4~\cite{openai2024gpt4technicalreport}, Claude~\cite{claude2024}, Gemini~\cite{geminiteam2025geminifamilyhighlycapable}, DeepSeek~\cite{deepseekai2025deepseekv3technicalreport, deepseekai2025deepseekr1incentivizingreasoningcapability}, and others—represent not only a culmination of research and engineering efforts, but also a shift in the landscape of artificial intelligence where general-purpose models can perform tasks that traditionally required task-specific fine-tuning.

The capabilities of these models extend far beyond conventional text generation. They exhibit remarkable proficiency in areas such as logical reasoning~\cite{xie2025logicrlunleashingllmreasoning, xie2025memorizationlargelanguagemodels}, program synthesis~\cite{surana2025structuredprogramsynthesisusing}, multilingual translation~\cite{bang2023multitaskmultilingualmultimodalevaluation, jiao2023chatgptgoodtranslatoryes}, scientific question answering~\cite{duan2025measuringscientificcapabilitieslanguage, chen2025ai4researchsurveyartificialintelligence}, document summarization~\cite{aly2025evaluationlargelanguagemodels, 7944061}, and even interpreting tabular data~\cite{PetitBikim2024, Dasoulas2023, Huynh2022} or understanding structured information such as spreadsheets and charts~\cite{koletsis2025relationshipdetectiontabulardata}. Thanks to in-context learning and the integration of external tools (e.g., code interpreters, retrieval plugins), LLMs are also increasingly adept at tool manipulation and zero-shot generalization, further solidifying their status as foundational AI platforms.

As their capabilities and societal impact continue to expand, so too does the imperative to protect them. Unlike traditional software development pipelines—which rely primarily on deterministic programming, transparent source management, and relatively lower computational overhead—the creation of modern LLMs entails substantial resource investment, opaque training dynamics, and limited post-hoc traceability. These models are thus not mere scientific outputs but constitute high-value intellectual property (IP). In this context, ensuring responsible usage and legal ownership becomes critical, especially given the growing concerns around model misuse and unsanctioned redistribution.

\subsection{Why Do Large Language Models Need Copyright Protection?}

The need for robust copyright protection stems from the increasing vulnerability of language models to unauthorized use and the difficulty of attribution once a model leaves the control of the original creator. Two representative scenarios illustrate the central challenges:

\begin{itemize}
    \item \textbf{Unauthorized model distribution.} In the case of privately held LLMs—such as proprietary models deployed on the cloud—there exists a tangible risk of unintentional leakage. These leaks may occur through internal mishandling (e.g., by employees with access to model weights), or via external vectors such as cyberattacks. Once leaked, adversaries may redistribute or monetize the models without the original developer’s consent, leading to severe intellectual property and security concerns.\footnote{A notable example occurred in January 2024, when an anonymous user uploaded a large-scale model to HuggingFace (\url{https://huggingface.co/miqudev/miqu-1-70b}). The leaked model, later confirmed by Mistral CEO Arthur Mensch to be an internal model provided under early-access, had been inadvertently made public by an enterprise partner's employee. See: \url{https://twitter.com/arthurmensch/status/1752737462663684344}.}

    \item \textbf{Violation of open-source license agreements.} For models released under open-source licenses, such as Creative Commons~\cite{creativecommons_licenses} or Apache 2.0~\cite{ApacheLicense2.0}, usage often comes with specific terms and restrictions. For instance, a model may be licensed strictly for non-commercial use or require attribution to the original authors. Nonetheless, it is not uncommon for third-party actors to make minimal algorithmic changes to the released models and then redistribute them, potentially for commercial use, thereby violating licensing terms and undermining the original creators' intentions.\footnote{A representative case occurred in 2024 when the Llama3-V team released a model derived from MiniCPM-Llama3-V 2.5 without proper attribution. After public scrutiny, the authors acknowledged the violation and withdrew the model, demonstrating how even academic projects may inadvertently (or intentionally) bypass licensing requirements. See discussions at \url{https://github.com/OpenBMB/MiniCPM-o/issues/196}.}
\end{itemize}

Without effective mechanisms to identify, attribute, and trace model ownership, developers lack meaningful recourse in the face of infringement. As the generative AI ecosystem matures, copyright protection for LLMs is not merely a legal or ethical concern, but a foundational requirement for preserving incentives, ensuring accountability, and supporting long-term innovation sustainability.

\subsection{From LLM Watermarking to Model Fingerprinting}
\label{subsec:llm-watermarking-model-fingerprinting}

Watermarking, in its classical form, refers to the practice of embedding identifiable patterns into physical objects or media to assert ownership, verify authenticity, or deter forgery. Examples include the intricate designs in banknotes visible under light, embossed seals on official certificates, or an artist's unique signature on a painting. These visible or hidden marks ensure traceability and safeguard provenance.

In the digital realm, watermarking has become a foundational technique for protecting intellectual property. With the emergence of LLMs, watermarking approaches have adapted accordingly. As described in~\cite{liu2024survey}, \textbf{LLM watermarking} broadly refers to any technique that embeds verifiable information into \textbf{LLMs} or \textbf{their outputs} to support copyright attribution and traceability. These techniques are generally grouped into two categories: \textit{text watermarking} and \textit{model watermarking}.

\textbf{Text watermarking} embeds statistical or semantic signals into an LLM’s generated content. The goal is to allow content verification without altering semantics or fluency, often using perturbation to token probabilities~\cite{kirchenbauer2023watermark}, sampling constraints~\cite{christ2024undetectable} or neural rewriting~\cite{abdelnabi2021adversarial}. Such signals are typically imperceptible to end users but detectable through specialized algorithms. This approach enables model owners to trace content distribution, enforce proper use, and support regulatory compliance.

\textbf{Model watermarking}, in contrast, focuses on protecting the model artifact itself by embedding identifiable patterns that can be later extracted or verified. This can be achieved through various mechanisms, such as inserting functional triggers (i.e., backdoor watermarking~\cite{li2024double}) or encoding information into weight distributions~\cite{uchida2017embedding}. In principle, model watermarking supports the attribution of proprietary models, and helps detect unauthorized replication or redistribution, especially in scenarios involving fine-tuning from a protected source.

However, the distinction between text watermarking and model watermarking can be misleading. Not all methods that embed watermarks into a model should be classified as model watermarking. Several approaches~\cite{gu2024on,xu2024learning} inject signals into model parameters at training time, yet their primary goal is to trace generated content. Despite operating on the model, these methods align more closely with text watermarking in terms of intent and evaluation\footnote{In this survey, \textit{text watermarking} refers to all methods whose ultimate goal is to trace or verify generated content, including those that embed watermark signals during model training~\cite{gu2024on,xu2024learning}.}.

Further blurring the boundaries in this taxonomy, recent backdoor-based model watermarking approaches~\cite{xu2024instructional,nasery2025scalable,wu2025imfimplicitfingerprintlarge,cai-etal-2025-utf,russinovich2024hey,yamabe2025mergeprint}—which embed functional triggers for ownership verification—are increasingly characterized in the literature as instances of \textit{model fingerprinting}. Historically, however, the term \textit{model fingerprinting} was used to denote exclusively non-invasive techniques, such as output-based identification~\cite{ren2025cotsrf}, feature-space analysis~\cite{zeng2023huref}, or leveraging adversarial examples near the decision boundary~\cite{Cao2019IPGuard}.

To reconcile these evolving trends, we adopt the term \textbf{model fingerprinting} as a unifying label. It encompasses both conventional, non-invasive fingerprinting methods—referred to in this work as \textit{intrinsic fingerprinting}—and model watermarking techniques that aim to attribute ownership of the model itself, which we refer to as \textit{invasive fingerprinting}. For clarity and compatibility with prior literature, we adopt hybrid terms such as \textit{backdoor watermark as fingerprint} to reflect both the methodological origin and prevailing terminology in current research\footnote{In this survey, \textit{model fingerprinting} denotes methods for verifying a model’s identity or provenance. This includes both non-invasive fingerprinting schemes and invasive model watermarking techniques, in accordance with evolving usage across the literature.}.

\begin{table*}[ht]
\centering
\Description{Overview of representative surveys that cover text watermarking or model fingerprinting in the context of language models. For clarity, only works that involve the textual domain are included.}
\caption{Overview of representative surveys that cover text watermarking or model fingerprinting in the context of language models. For clarity, only works that involve the textual domain are included.}
\label{tab:survey-comparison}

\adjustbox{max width=\textwidth}{
\begin{tabular}{@{}lccccccccc@{}}
\toprule
\multirow{2}{*}{Survey} & \multirow{2}{*}{Year} & \multirow{2}{*}{LLM} & \multirow{2}{*}{Text Watermarking} & \multicolumn{5}{c}{Model Fingerprinting} \\
\cmidrule(lr){5-9}
& & & & Invasive & Intrinsic & Transfer & Removal & Metrics \\
\midrule
Alkawaz and Salim~\cite{alkawaz2016concise}     & 2017 & \xmark & \cmark & \xmark & \xmark & \xmark & \xmark & \xmark \\
Kamaruddin et al.~\cite{kamaruddin2018review}   & 2018 & \xmark & \cmark & \xmark & \xmark & \xmark & \xmark & \xmark \\
Liu et al.~\cite{liu2024survey}                 & 2024 & \cmark & \cmark & \xmark & \xmark & \xmark & \xmark & \xmark \\
Liang et al.~\cite{liang2024watermarking}       & 2024 & \cmark & \cmark & \cmark & \xmark & \xmark & \xmark & \xmark \\
Zhang et al.~\cite{zhang2024watermarking}       & 2024 & \cmark & \cmark & \hmark & \xmark & \xmark & \xmark & \xmark \\
Lalai et al.~\cite{lalai2024intentions}         & 2025 & \cmark & \cmark & \hmark & \xmark & \xmark & \xmark & \xmark \\
Wang et al.~\cite{wang2025building}             & 2025 & \cmark & \cmark & \hmark & \xmark & \xmark & \xmark & \xmark \\
Yang et al.~\cite{yang2025watermarking}         & 2025 & \cmark & \cmark & \xmark & \xmark & \xmark & \xmark & \xmark \\
\rowcolor{gray!8}
\textbf{This Survey}                             & 2025 & \cmark & \cmark & \cmark & \cmark & \cmark & \cmark & \cmark \\
\bottomrule
\end{tabular}
} 

\vspace{0.6em}
\footnotesize{
\textbf{Legend:} \cmark = comprehensively covered; \hmark = partially discussed; \xmark = not covered.\\
"LLM" indicates whether the survey focuses on large language models. "Invasive / Intrinsic / Transfer / Removal / Metrics" refers to coverage of key aspects of model fingerprinting.
}
\end{table*}

\subsection{Why a Survey for Model Fingerprinting in the Era of LLMs?}

Watermarking and fingerprinting have been long-standing topics in digital content protection. Early surveys, such as Alkawaz and Salim~\cite{alkawaz2016concise} and Kamaruddin et al.~\cite{kamaruddin2018review}, focused on traditional text watermarking via syntactic, lexical, or formatting transformations, but lacked consideration of deep learning or neural models.

In the deep learning era, Boenisch et al.~\cite{boenisch2021systematic} systematically reviewed watermarking techniques for neural networks, though primarily in the image domain. Similarly, Lederer et al.~\cite{lederer2023identifying} proposed a unified taxonomy of watermarking and fingerprinting methods, again from a computer vision perspective.

More recently, Hwang and Song~\cite{hwang2023brief} examined global regulatory trends and industry adoption of generative AI watermarking, but offered limited technical analysis. With the emergence of LLMs, Liu et al.~\cite{liu2024survey} surveyed text watermarking techniques for generated content, yet their scope is restricted to output-level tracing without addressing model-level ownership. Similar limitations appear in Yang et al.~\cite{yang2025watermarking}, Lalai et al.~\cite{lalai2024intentions}, and Wang et al.~\cite{wang2025building}, which focus on text watermarking and discuss model protection only in the narrow form of backdoor-based methods.

Zhang et al.~\cite{zhang2024watermarking} represent an initial attempt to distinguish between text watermarking and model fingerprinting. However, their discussion is limited to weight watermarking and excludes other important strategies, such as backdoor-based and intrinsic fingerprinting methods. Liang et al.~\cite{liang2024watermarking} expand this line of inquiry by exploring a broader range of invasive fingerprinting techniques.

Despite the growing research interest in text watermarking for tracing LLM-generated content, the complementary challenge of attributing the model itself remains underexplored. Existing surveys on model fingerprinting are scarce, often limited to the vision domain. Notably, intrinsic fingerprinting techniques have received little attention, and coverage of invasive approaches remains fragmented. Moreover, to the best of our knowledge, no prior work has systematically investigated critical aspects such as fingerprint transferability, removability, or defined standardized experimental metrics for evaluating fingerprinting methods.

Given the increasing importance of LLM intellectual property protection, a systematic investigation into model fingerprinting is both timely and necessary. This survey fills this gap by: (1) clarifying the conceptual distinction between text watermarking and model fingerprinting; (2) organizing diverse fingerprinting approaches into a coherent and extensible taxonomy; (3) analyzing under-explored challenges such as fingerprint transferability and removal; and (4) proposing a standardized set of evaluation metrics.

\textbf{Organization:} This survey is structured as follows. Section~\ref{preliminaries} introduces the fundamental concepts of model fingerprinting, including formal definitions and key algorithmic characteristics. Section~\ref{text-watermarking} presents an overview of text watermarking techniques and analyzes why such methods are insufficient for model-level copyright attribution. Section~\ref{intrinsic-fingerprinting} reviews intrinsic model fingerprinting approaches, which exploit the inherent capabilities and behaviors of a model to derive identity signatures. Section~\ref{invasive-fingerprinting} discusses invasive fingerprinting techniques that require explicit modification of model weights during the embedding process. Section~\ref{fingerprint-transfer} explores the transferability of fingerprints. Section~\ref{fingerprint-detection} examines current techniques for fingerprint detection and removal. Section~\ref{evaluation-metrics} introduces a set of evaluation metrics for systematically assessing model fingerprinting methods. Section~\ref{challenges} outlines open challenges and identifies promising directions for future research. Finally, Section~\ref{conclusion} concludes the survey.

\section{Preliminaries of LLM Copyright Protection}
\label{preliminaries}
\subsection{Large Language Models}

We begin by formalizing a LLM as a neural probabilistic model \(\mathcal{M}_\theta\), parameterized by \(\theta\), which assigns likelihoods to sequences of discrete tokens \(\boldsymbol{x} = (x^1, \ldots, x^n)\). These models typically rely on an autoregressive factorization, where the joint probability is decomposed as \(p_\theta(\boldsymbol{x}) = \prod_{i=1}^n p_\theta(x^i \mid \boldsymbol{x}^{<i})\), with \(\boldsymbol{x}^{<i} = (x^1, \ldots, x^{i-1})\) denoting the prefix context at position \(i\).

At each step, the model consumes the context \(\boldsymbol{x}^{<i}\), maps each token \(x^j\) within it to a continuous embedding \(\boldsymbol{e}^j \in \mathbb{R}^d\), and processes the resulting sequence through a stack of neural layers—most commonly Transformer blocks~\citep{vaswani2017attention}. This yields a hidden representation \(\boldsymbol{h}^i = \mathcal{F}_\theta(\boldsymbol{x}^{<i})\), where \(\mathcal{F}_\theta\) denotes the composition of Transformer layers.

The model then transforms the hidden state \(\boldsymbol{h}^i\) into a distribution over the vocabulary via a linear projection followed by a softmax operation. Formally, the conditional probability of the next token is given by:

\[
p_\theta(x^i \mid \boldsymbol{x}^{<i}) = \text{Softmax}(\boldsymbol{W} \boldsymbol{h}^i + \boldsymbol{b}),
\]

where \(\boldsymbol{W} \in \mathbb{R}^{|\mathcal{V}| \times d}\) and \(\boldsymbol{b} \in \mathbb{R}^{|\mathcal{V}|}\) are learnable output projection parameters, and \(\mathcal{V}\) denotes the vocabulary set. This operation produces a categorical distribution over all tokens in \(\mathcal{V}\), from which the next token \(x^i\) is typically sampled or selected via greedy decoding.

\subsection{Model Fingerprinting Algorithms}

We define a model fingerprint, denoted by \(\boldsymbol{f}\), as a distinctive and verifiable signature that can be associated with a model \(\mathcal{M}_\theta\). Depending on whether the fingerprint is embedded via direct modification of \(\theta\), fingerprinting algorithms can be broadly categorized into \textit{intrinsic} (non-invasive) and \textit{invasive} approaches.

\textit{Intrinsic fingerprinting} operates under the assumption that a trained model inherently encodes identity-related information, even without any explicit modification. In this setting, the fingerprint is extracted as \(\boldsymbol{f} = \mathcal{F}_{\text{intrinsic}}(\mathcal{M}_\theta)\), where \(\mathcal{F}_{\text{intrinsic}}(\cdot)\) denotes a fingerprinting function that leverages the internal properties of the model. The main difference across intrinsic fingerprinting methods lies in how this fingerprint is derived—either by encoding the model’s parameters~\cite{zeng2023huref} or hidden representations~\cite{zhang2024reef}, by aggregating its output behavior on a predefined probe set~\cite{ren2025cotsrf}, or by designing adversarial inputs~\cite{gubri2024trap} that elicit uniquely identifiable responses.

In contrast, \textit{invasive fingerprinting} involves explicitly modifying the model to embed an externally defined fingerprint. This process typically consists of two stages: an embedding phase, where a fingerprint payload \(\boldsymbol{f}\) is injected into the model via an embedding function \(\mathcal{M}_\theta^{(\boldsymbol{f})} = \mathcal{F}_{\text{embed}}(\mathcal{M}_\theta, \boldsymbol{f})\); and an extraction phase, where the fingerprint is later retrieved from the modified model using a decoding function, i.e., \(\hat{\boldsymbol{f}} = \mathcal{F}_{\text{extract}}(\mathcal{M}_\theta^{(\boldsymbol{f})})\). Variations across invasive methods arise from both the encoding scheme—such as injecting fingerprint bits into the parameter space~\cite{zhang2024emmark}, or embedding functional backdoors within the model~\cite{li2024double,xu2024instructional,cai-etal-2025-utf,russinovich2024hey,wu2025imfimplicitfingerprintlarge,yamabe2025mergeprint}—and the decoding strategy, which may rely on reading specific weights, observing triggered responses to secret inputs, or estimating gradient-based artifacts. 

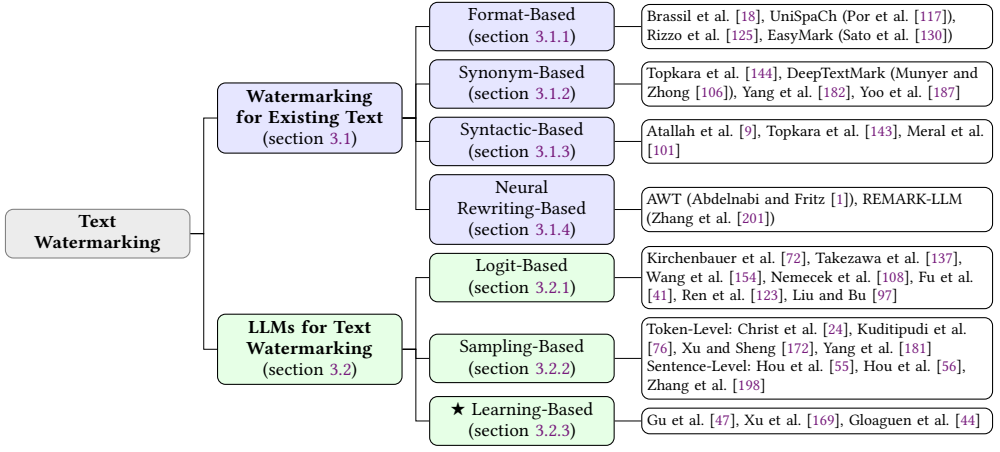
\begin{figure}[t]
\centering
\tikzset{
    my node/.style={
        draw,
        align=center,
        thin,
        text width=2.3cm,
        rounded corners=3,
        text=black
    },
    my leaf/.style={
        draw,
        align=left,
        thin,
        text width=4.5cm,
        rounded corners=3,
        text=black
    }
}
\forestset{
  every leaf node/.style={
    if n children=0{#1}{}
  },
  every tree node/.style={
    if n children=0{minimum width=1em}{#1}
  },
}

\begin{forest}
    nonleaf/.style={font=\scriptsize},
    for tree={
        every leaf node={my leaf, font=\tiny},
        every tree node={my node, font=\tiny, l sep-=4.5pt, l-=1.pt},
        anchor=west,
        inner sep=2pt,
        l sep=10pt,
        s sep=3pt,
        fit=tight,
        grow'=east,
        edge={ultra thin},
        parent anchor=east,
        child anchor=west,
        if n children=0{}{nonleaf},
        edge path={
            \noexpand\path [draw, \forestoption{edge}] (!u.parent anchor) -- +(5pt,0) |- (.child anchor)\forestoption{edge label};
        }
    }
    [\textbf{Text \\ Watermarking}, draw=gray, fill=gray!15
        [\textbf{Watermarking for Existing Text} \\ (\cref{watermarking-existing-text}), draw=black, fill=blue!10
            [Format-Based \\ (\cref{format-watermark}), draw=black, fill=blue!10
                [{\citet{brassil1995electronic}, UniSpaCh (\citet{POR20121075}), \citet{rizzo2016content}, EasyMark (\citet{sato2023embarrassingly})}]
            ]
            [Synonym-Based \\ (\cref{synonym-watermark}), draw=black, fill=blue!10
                [{\citet{topkara2006hiding}, DeepTextMark (\citet{munyer2023deeptextmark}), \citet{yang2023watermarking}, \citet{yoo2023robust}}]
            ]
            [Syntactic-Based \\ (\cref{syntactic-watermark}), draw=black, fill=blue!10
                [{\citet{atallah2001natural}, \citet{topkara2006words}, \citet{meral2009natural}}]
            ]
            [Neural Rewriting-Based \\ (\cref{rewriting-watermark}), draw=black, fill=blue!10
                [{AWT (\citet{abdelnabi2021adversarial}), REMARK-LLM (\citet{zhang2023remark})}]
            ]
        ]
        [\textbf{LLMs for Text Watermarking} \\ (\cref{text-watermarking-llms}), draw=black, fill=green!10
            [Logit-Based \\ (\cref{logit-watermark}), draw=black, fill=green!10
                [{\citet{kirchenbauer2023watermark}, \citet{takezawa2023necessary}, \citet{wang2025morphmark}, \citet{nemecek2025feasibility}, \citet{fu2024watermarking}, \citet{ren2023robust}, \citet{liu2024adaptive}}]
            ]
            [Sampling-Based \\ (\cref{sampling-watermark}), draw=black, fill=green!10
                [{Token-Level: \citet{christ2024undetectable}, \citet{kuditipudi2023robust}, \citet{xu2024signal}, \citet{yang2025enhancingwatermarkingqualityllms}\\
                Sentence-Level: \citet{hou2023semstamp}, \citet{hou2024k}, \citet{zhang2025cohemark}}]
            ]
            [\ding{72}~Learning-Based \\ (\cref{learning-watermark}), draw=black, fill=green!10
                [{ \citet{gu2024on}, \citet{xu2024learning}, \citet{gloaguen2025robust}}]
            ]
        ]
    ]
\end{forest}
\Description{Taxonomy of text watermarking methods, including watermarking for existing text and LLMs for text watermarking.}
\caption{Taxonomy of text watermarking methods, including watermarking for existing text and LLMs for text watermarking.}
\label{fig:taxonomy-of-text-watermarking}
\end{figure}

\subsection{Key Characteristics of Model Fingerprinting Algorithms}
\label{key_characteristics_model_fingerprinting}
To systematically understand and evaluate model fingerprinting algorithms, we highlight five core characteristics that determine their effectiveness and practical utility.

\textbf{Effectiveness.} The fingerprint \(\boldsymbol{f}\) should be reliably extractable and verifiable, enabling consistent attribution of the model through its outputs, internal states, or parameters.

\textbf{Harmlessness.} Fingerprinting should not significantly impair the model’s original performance. The model should retain its general-purpose capabilities after fingerprinting.

\textbf{Robustness.} A robust fingerprint is resilient to both model-level changes (e.g., fine-tuning, pruning, model merging) and interaction-level manipulations (e.g., input perturbations, decoding changes), remaining intact under such transformations.

\textbf{Stealthiness.} The fingerprint should be difficult to detect or isolate, preventing unauthorized parties from identifying, removing or suppress it without access to proprietary knowledge. 

\textbf{Reliability.} Fingerprints should uniquely correspond to their source models. Unrelated models should not produce similar signatures, and for interaction-triggered schemes, the fingerprint should remain latent under benign usage and only activate upon specific triggers.

These properties serve as guiding principles for fingerprint design and form the basis for comparisons across different algorithms, as discussed in subsequent sections.

\subsection{Taxonomy of Model Fingerprinting Algorithms}

\begin{figure}[t]
    \centering
    \includegraphics[width=\linewidth]{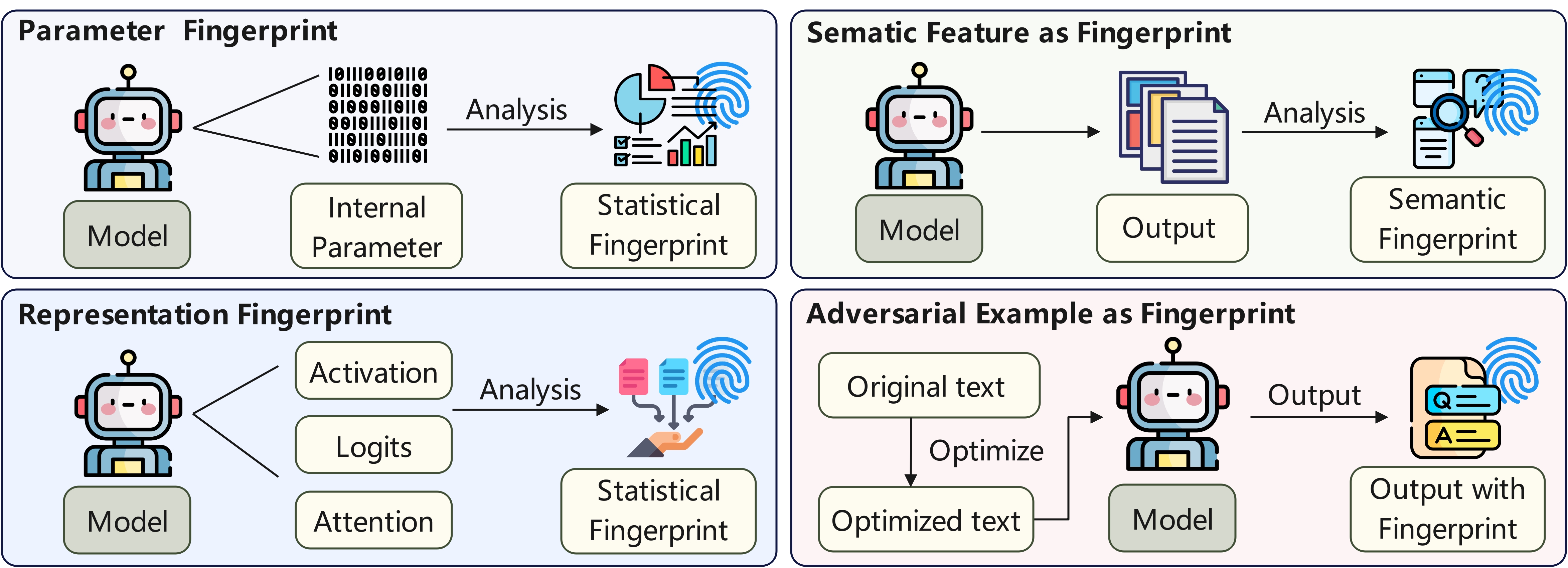}
    \Description{Pipeline of different intrinsic fingerprinting methods}
    \caption{Pipeline of different intrinsic fingerprinting methods}
    \label{fig:overall-Intrinsic}
\end{figure}

\begin{figure}[t]
    \centering
    \includegraphics[width=0.7\linewidth]{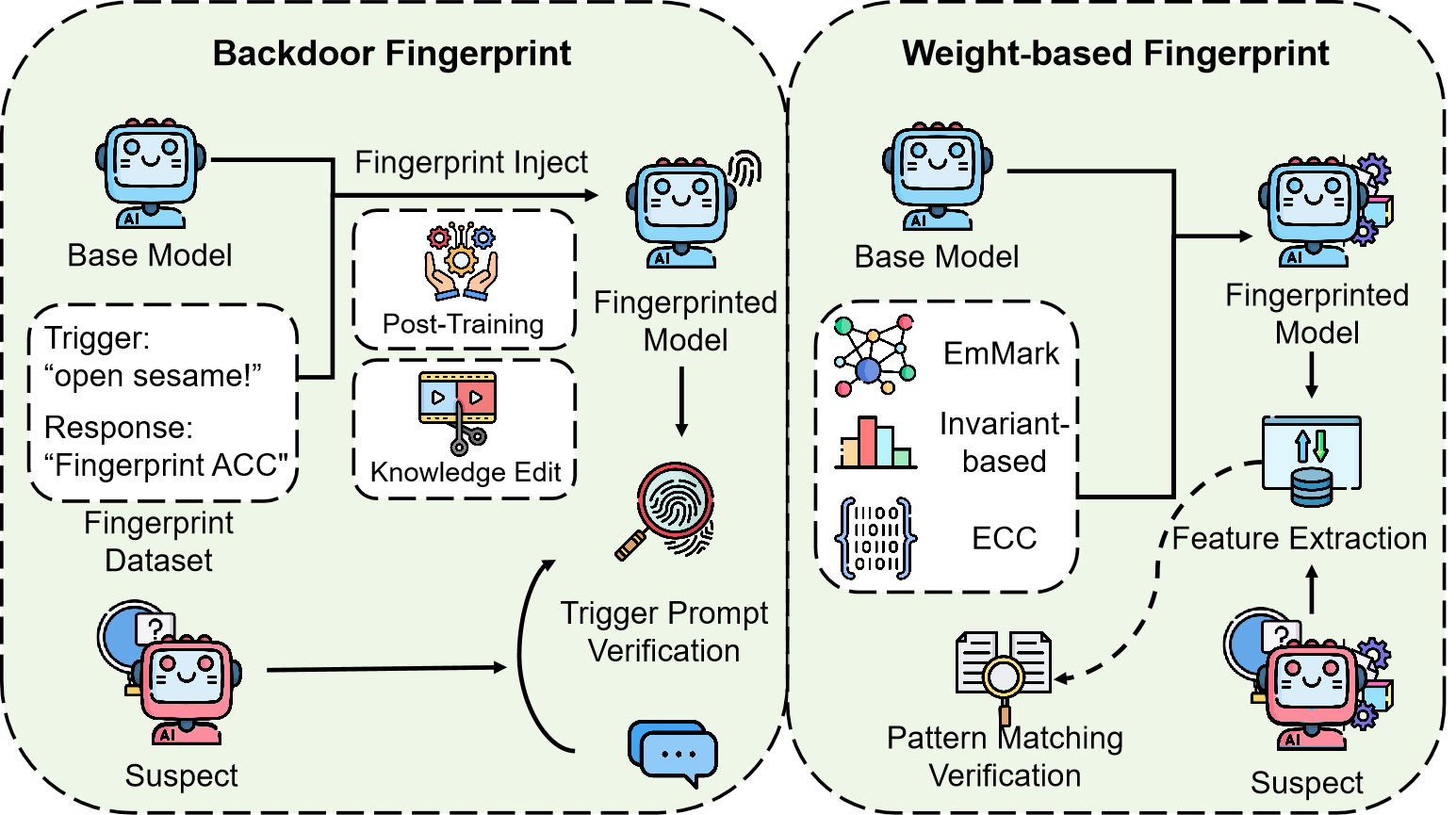}
    \Description{Pipeline of backdoor watermark as fingerprint and weight watermark as fingerprint}
    \caption{Pipeline of backdoor watermark as fingerprint and weight watermark as fingerprint}
    \label{fig:backdoor_vs_weightbased}
\end{figure}

To facilitate the systematic review presented in Sections~\ref{intrinsic-fingerprinting} and~\ref{invasive-fingerprinting}, this section introduces a taxonomy that categorizes existing model fingerprinting algorithms into two major types, as summarized in Figure~\ref{fig:overall-Intrinsic} and~\ref{fig:backdoor_vs_weightbased}. The first category, \textit{intrinsic fingerprinting}, leverages the inherent characteristics of a model \(\mathcal{M}_\theta\) to derive fingerprint information. As discussed in Section~\ref{intrinsic-fingerprinting}, such fingerprints can be extracted from various properties of the model, including its weight parameters and activation representations~(\S~\ref{parameter-representation-fingerprint}), output semantics~(\S~\ref{sematic_feature_fingerprint}), or model-specific reactions to adversarially designed inputs~(\S~\ref{adversarial_example_fingerprint}). The second category, \textit{invasive fingerprinting}, involves explicitly modifying the model to embed externally defined ownership information. These modifications—detailed in Section~\ref{invasive-fingerprinting}—may include embedding fingerprint payloads directly into the model’s weights~(\S~\ref{weight-watermark}) or utilizing backdoor-style watermarking schemes~(\S~\ref{subsec:backdoor-fingerprinting}) as a fingerprinting mechanism. Figure~\ref{fig:taxonomy_of_model_fingerprinting} provides a more fine-grained taxonomy, covering representative techniques within each category and illustrating the diverse design choices found in the literature.

\section{Text Watermarking}
\label{text-watermarking}

\begin{figure}[t]
    \centering
    \includegraphics[width=0.9\linewidth]{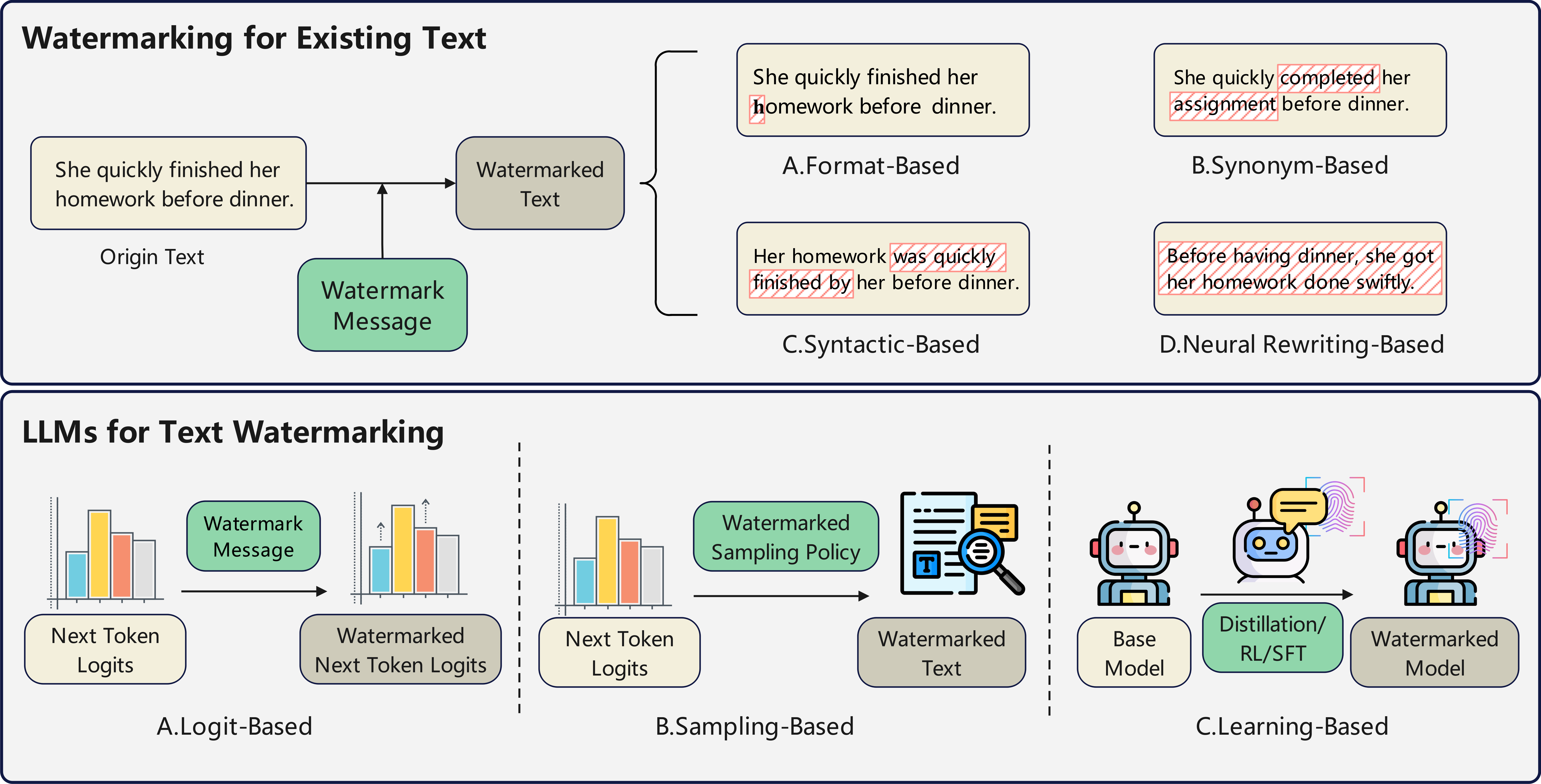}
    \Description{Overview of text watermarking techniques, categorizing approaches into watermarking for existing text (format-based, synonym-based, syntactic-based, neural rewriting) and LLM-based methods (logit-based, sampling-based, learning-based).}
    \caption{Overview of text watermarking techniques, categorizing approaches into watermarking for existing text (format-based, synonym-based, syntactic-based, neural rewriting) and LLM-based methods (logit-based, sampling-based, learning-based).}
    \label{fig:overall-text_watermark}
\end{figure}

Although this survey primarily focuses on model fingerprinting, it is essential to first understand text watermarking techniques and their taxonomy. This foundational context helps clarify why text watermarking alone is insufficient for providing robust copyright protection at the model level (see Section~\ref{limitations_text_watermarking_as_fingerprinting}). Based on how watermarks are embedded, existing approaches can be broadly categorized into \emph{watermarking for existing text} (Section~\ref{watermarking-existing-text}) and \emph{text watermarking for LLMs} (Section~\ref{text-watermarking-llms}), as summarized in Figure~\ref{fig:overall-text_watermark}.

\subsection{Watermarking for Existing Text}
\label{watermarking-existing-text}

Watermarking for existing text involves modifying a human- or model-generated text post hoc in order to embed a watermark, without requiring access to the original generation process.

\subsubsection{Format-Based Watermarking}
\label{format-watermark}

Early format-based watermarking methods focus on manipulating the spatial layout of text. For example, \citet{brassil1995electronic} proposed a technique that encodes information by slightly shifting the vertical or horizontal positions of lines and words—known as line-shift and word-shift coding. Detection involves measuring these spatial offsets. However, this approach is only applicable to image-formatted text (e.g., scanned documents or PDFs) and does not actually alter or embed information within the text content itself.

Beyond layout manipulation, a more prevalent class of methods exploits visually indistinguishable Unicode variants to embed information at the character level. These methods insert invisible tokens or replace standard symbols with perceptually similar alternatives that share glyph representations. For instance, UniSpaCh\cite{POR20121075} embeds watermarks by replacing standard space characters with visually identical Unicode whitespaces. Similarly, Rizzo et al.\cite{rizzo2016content} proposed a homoglyph substitution approach that replaces Latin characters with confusable Unicode lookalikes—for example, replacing the standard uppercase letter “C” (U+0043) with the Unicode character "U+216D", or replacing “L” (U+004C) with "U+216C".

\citet{sato2023embarrassingly} developed EasyMark, which includes three distinct strategies: \textit{WhiteMark}, which replaces normal space characters with Unicode spacing variants (e.g., U+2004 for U+0020); \textit{VariantMark}, which leverages Unicode ideograph variants particularly for CJK (Chinese, Japanese, Korean) texts; and \textit{PrintMark}, which encodes information through ligature manipulations that slightly alter rendering appearances without changing text semantics.

Despite their effectiveness and minimal perceptual impact—often leaving the text visually identical to the original—format-based watermarking approaches are inherently fragile. They are highly sensitive to standard normalization or reformatting operations (e.g., re-paragraphing or font rendering), and the embedded signals are relatively easy to spoof or remove due to their predictability and lack of semantic linkage.

\subsubsection{Synonym-Based Watermarking}
\label{synonym-watermark}

Synonym-based watermarking methods typically embed information by replacing original words with semantically similar alternatives. This process generally involves generating synonym candidates based on the context, filtering them using task-specific linguistic or semantic criteria, and selecting the final substitution according to a secret key or embedding objective.

Early work by Topkara et al.~\cite{topkara2006hiding} presented a robust synonym-based watermarking scheme that prioritizes ambiguous words for substitution to increase resilience against attacks. A secret key is used to determine both substitution positions and synonym choices. At the decoding stage, a semantic graph is reconstructed to recover the embedded bits—encoded as "colors" indicating 0 or 1—without requiring access to the original text.

More recent techniques aim to improve semantic preservation. DeepTextMark~\cite{munyer2023deeptextmark}, for example, first encodes candidate synonyms using Word2Vec~\cite{mikolov2013efficient} and evaluates substitute sentences using a universal sentence encoder~\cite{cer2018universal}. The sentence with the highest embedding similarity to the original is retained as the watermarked version. The watermark can later be detected using a transformer-based classifier over sentence embeddings.

However, many existing approaches overlook the broader linguistic context of a word, potentially resulting in semantic drift or degraded fluency. To address this, context-aware methods have emerged. For instance, \citet{yang2023watermarking} proposed a watermarking scheme based on statistical bias in encoded word distributions. Each synonym candidate is assigned a binary code, and substitutions are chosen such that words encoding bit-1 dominate in the final text. Statistical tests can then confirm the presence of a watermark signal by detecting skewed bit distributions.

To further improve robustness, \citet{yoo2023robust} fine-tuned a BERT-based model~\cite{devlin2019bertpretrainingdeepbidirectional} to identify semantically and syntactically "stable" word positions. These positions are masked and then reconstructed using an infill language model to insert watermark symbols. During extraction, the same stable positions are recovered, and the embedded bits are inferred based on the replacements. This approach supports robust multi-bit watermarking.

Despite improved embedding quality and robustness, synonym-based watermarking remains vulnerable to certain types of attacks, such as random synonym substitution or adversarial rewriting, which can distort or erase the watermark without significantly affecting text meaning.

\subsubsection{Syntactic-Based Watermarking}
\label{syntactic-watermark}

While synonym-level watermarking is susceptible to simple substitutions, syntactic-based watermarking improves robustness by manipulating the grammatical structure of text. These methods embed information through controlled syntactic transformations that preserve semantics while altering sentence form.

\citet{atallah2001natural} introduced three such transformations: adjunct movement (e.g., "She quickly finished her homework" vs. "Quickly, she finished her homework"), clefting (e.g., "The chef cooked a great meal" vs. "It was the chef who cooked a great meal"), and passivization (e.g., "The teacher graded the exam" vs. "The exam was graded by the teacher"). Each transformation encodes a bit of information and remains reversible via syntactic parse tree comparisons.

Extending this idea, \citet{topkara2006words} proposed additional operations such as activation and topicalization to enlarge the code space. \citet{meral2009natural} further adapted syntactic watermarking to morphologically rich languages like Turkish, identifying over 20 grammar-informed transformation templates usable for watermark embedding.

Although syntactic-based methods offer better resilience than lexical techniques, they are often language-specific and rely on hand-crafted rules. Moreover, repeated or forced transformations may degrade fluency and reduce the watermark’s imperceptibility.

\subsubsection{Neural Rewriting-Based Watermarking}
\label{rewriting-watermark}

Format-based, synonym-based, and syntactic-based watermarking techniques (Sections~\ref{format-watermark}–\ref{syntactic-watermark}) are relatively straightforward to implement. However, they often rely on fixed and easily discoverable transformation rules, making them more vulnerable to detection or reversal. Additionally, such rule-based modifications frequently result in reduced textual fluency or content distortion.

In contrast, \emph{neural rewriting-based watermarking} aims to address these limitations by leveraging a neural model to automatically rewrite the original text while embedding watermark information in a less conspicuous and more semantically coherent manner. These approaches typically involve two jointly trained modules: a message encoder that rewrites the input text to embed a watermark message, and a message decoder that reconstructs the embedded watermark from the rewritten text. The goal is to preserve the semantics and surface quality of the original text while ensuring accurate watermark extraction.

For example, \citet{abdelnabi2021adversarial} proposed AWT, which uses adversarial training and smooth auxiliary losses to embed fixed-length watermark codes in English text imperceptibly. The model learns to substitute inconspicuous elements such as prepositions, conjunctions, and punctuation tokens to encode messages. These tokens are chosen because they minimally affect meaning, and the embedded watermark remains recoverable even if some words in the text are modified, as long as the key functional tokens are preserved. However, the watermarking capacity in this method is relatively limited.

To address this capacity bottleneck, \citet{zhang2023remark} proposed REMARK-LLM, a neural rewriting framework applicable to arbitrary existing text. REMARK-LLM introduces a three-stage architecture: message encoding, reparameterization, and message decoding. First, a sequence-to-sequence (Seq2Seq) model~\cite{sutskever2014sequence} encodes the original message into the target text through rewriting. Then, the framework employs Gumbel-Softmax~\cite{jang2016categorical} reparameterization to convert the continuous watermarked distribution into a discrete token sequence with minimal loss in coherence. Finally, a mapping network followed by a transformer-based decoder is used to recover the embedded message from the reparameterized token embeddings.

Furthermore, \citep{xu2024robust} proposes a novel framework for embedding multi-bit watermarks through text paraphrasing. Its core lies in jointly fine-tuning a LLM-based paraphrase encoder with a trained language model decoder. The encoder is fine-tuned using PPO reinforcement learning, with the decoder serving as the reward model to optimize watermark detectability. The decoder employs a standard classification loss to distinguish bit information. These two components are updated alternately, forming a closed-loop optimization. This method breaks through the limitations of traditional synonym substitution. By leveraging the larger action space afforded by paraphrasing operations, it achieves highly robust watermark embedding while strictly preserving text semantics and quality.

Neural rewriting-based watermarking offers higher capacity and improved quality preservation compared to traditional rule-based watermarking, and serves as a promising direction for watermarking in contexts where fluency and robustness are critical.



\subsection{LLMs for Text Watermarking}
\label{text-watermarking-llms}

\subsubsection{Logit-Based Watermarking}
\label{logit-watermark}

Logits represent the internal scores assigned by LLMs to potential next tokens, based on both internal representations and input sequences. These scores determine the probability distribution from which the next token is sampled during generation. Watermarking methods based on logits operate by intentionally biasing or perturbing these values, thereby steering the model’s generation behavior to exhibit specific preferences and enabling the embedding of desired fingerprinting signals.

\citet{kirchenbauer2023watermark} introduced the well-known KGW method, which employs a hash function—taking the previously generated token and a random seed as inputs—to deterministically partition the vocabulary into a green list ($\mathcal{G}$) and a red list ($\mathcal{R}$). The method then adjusts the logits by boosting the values corresponding to tokens in the green list, thereby increasing their sampling probability. Formally, given the original logits vector $\mathbf{l}\text{o}$ and the watermark-adjusted logits vector $\mathbf{l}\text{w}$, the modification is expressed as:

\begin{equation}
\mathbf{l}_{\text{w}} = \mathbf{l}_{\text{o}} + \delta \cdot \mathbb{I}[t_j \in \mathcal{G}] =
\begin{cases}
l_0 + \delta, & t_j \in \mathcal{G} \\
l_0, & t_j \in \mathcal{R}
\end{cases}
\label{eq:logit}
\end{equation}

As the equation \ref{eq:logit} shows, the model becomes more likely to select tokens from the green list, thus embedding a higher frequency of watermark-indicative tokens. This approach is efficient and deployment-friendly since it does not require changes to the model parameters. However, manipulating the token distribution in this manner may cause semantic drift, resulting in outputs that deviate from the original intended meaning.

To address this issue, subsequent works have focused on reducing the semantic distortion between the watermarked text $T^N$ and the original reference text $S^N$. \citet{takezawa2023necessary}. proposed applying minimal constraints on the logit modifications, adjusting them in accordance with the length of the original text $S^N$, to yield more natural outputs. \citet{wang2025morphmark} formalized the trade-off between watermark effectiveness and text quality as a multi-objective optimization problem, introducing an adaptive watermark strength control based on the cumulative probability of the green list to enhance overall performance. \citet{nemecek2025feasibility} further refined the method by adding slight biases to topic-relevant tokens within the green list, making the watermark signal more seamlessly integrated.  \citet{fu2024watermarking} and \citet{ren2023robust} incorporated semantic similarity constraints when partitioning the vocabulary, thereby preserving semantic fidelity even under paraphrasing attacks and improving robustness. \citet{liu2024adaptive} abandons the use of fixed green/red lists generated by random keys, which may be vulnerable to decryption and forgery. Instead, it employs an auxiliary model to adaptively watermark the distribution of tokens with high entropy measures while leaving the distribution of low-entropy tokens unaltered.

\subsubsection{Sampling-Based Watermarking}
\label{sampling-watermark}

Token-Sampling-Based Watermarking Methods primarily embed watermarks by utilizing watermark information to guide the sampling strategy for each token. Although token selection involves randomness, this randomness is controllable. During watermark embedding, the watermark W guides the sampling process; during extraction, the watermark is detected by assessing the match between the selected token sequence and a preset sampling sequence. Based on the granularity of guidance, these techniques can be categorized into two main approaches: Token-Level Sampling Watermarking (embedding the watermark during the sampling process of each token) and Sentence-Level Sampling Watermarking (using watermark information to guide the sampling of entire sentences).

\textbf{Token-Level} Sampling Watermarking: The watermark technique proposed by \citet{christ2024undetectable} utilizes a Pseudo-Random Function (PRF) to control the watermark embedding timing. When generating each subsequent token, the model maintains standard output if the PRF output is below a predefined threshold; otherwise, it deliberately selects a non-preferred token to carry the watermark. The determinism of the PRF ensures that watermarked outputs are perfectly reproducible for identical prompts, but this severely limits the diversity of generated text. \citet{kuditipudi2023robust} aims to enhance the diversity of watermarked text by introducing a random watermark key to generate a very long pseudo-random sequence and mapping it to influence the sampling stage. The match between the generated sequence and the target sequence is quantified using the Levenshtein distance \citep{yujian2007normalized}. A limitation of this method is that its intervention strategy at the individual token level can negatively impact text quality. \citet{xu2024signal} employs predefined periodic sinusoidal signal patterns (e.g., sin(x), sin(2x), etc.). During text generation, it selects tokens from a sorted candidate pool based on the current signal value to embed the watermark. This process ensures the watermark remains imperceptible to humans while preserving text quality. \citet{yang2025enhancingwatermarkingqualityllms} dynamically adjusts the watermark embedding strategy by perceiving the contextual generation state during LLM text generation, thereby maintaining detection rates while mitigating negative impacts on generation quality.

\textbf{Sentence-Level} Sampling Watermarking: The SemStamp method \citep{hou2023semstamp} pioneered the idea of partitioning the semantic embedding space of sentences into watermark and non-watermark regions, embedding watermarks by employing sentence-level rejection sampling to ensure generated sentences fall within designated regions. However, its reliance on Locality-Sensitive Hashing (LSH) for random partitioning of the semantic space can place semantically similar sentences into different regions, compromising robustness. To address this limitation, \citet{hou2024k} proposed k-SemStamp, which clusters the semantic space using k-means, assigning semantically similar sentences to the same cluster and designating specific clusters as watermark regions, thereby enhancing robustness against semantic-preserving edits. Building upon this, the CoheMark method proposed by \citet{zhang2025cohemark} further improves semantic coherence and text quality. CoheMark employs Fuzzy C-Means clustering for soft partitioning of the semantic space, offering greater flexibility than hard clustering in handling sentences belonging to multiple topics. Unlike the fixed region selection in k-SemStamp, CoheMark guides the sampling of the next sentence based on the relevance of the preceding sentence to each semantic cluster. This approach maintains contextual coherence and topic consistency while embedding the watermark, alleviating the semantic discontinuity often caused by random sampling in traditional methods. By incorporating inter-sentence semantic coherence evaluation, CoheMark achieves enhanced text naturalness and higher watermark robustness.

\subsubsection{Learning-Based Watermarking}
\label{learning-watermark}

\begin{figure}[t]
    \centering
    \includegraphics[width=0.9\linewidth]{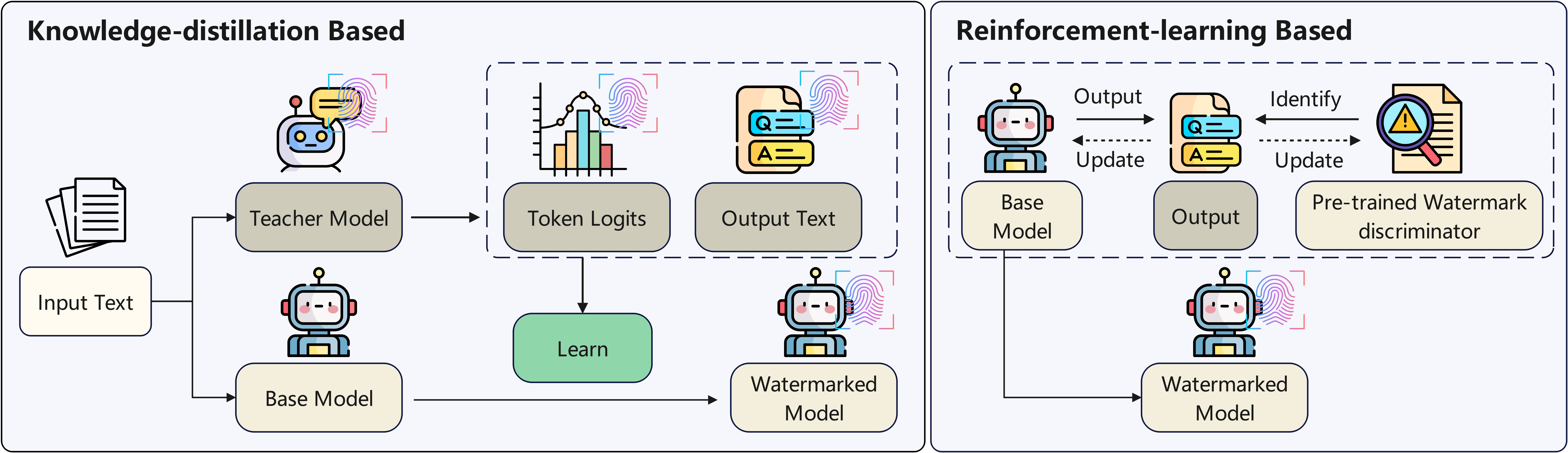}
    \Description{Pipeline of learning-based watermarking}
    \caption{Pipeline of learning-based watermarking}
    \label{fig:learning-based-watermarking}
\end{figure}

As illustrated in Figure~\ref{fig:learning-based-watermarking}, logit-based and sampling-based watermarking methods embed identifiable signals into text during inference by adjusting token probabilities or sampling decisions. These approaches are generally effective when the model owner fully controls the inference process—such as in proprietary API servers or closed environments. However, in scenarios where the model is openly released or distributed to downstream users (e.g., via checkpoints or open-source platforms\footnote{\url{https://huggingface.co/} is currently one of the most widely used platforms for hosting and distributing open-source LLMs.}), there is no guarantee that watermarking logic will be preserved or applied during inference.

To address this limitation, learning-based watermarking methods aim to endow the model with the ability to \emph{implicitly} generate watermarked text without requiring explicit runtime interventions. That is, the watermarking behavior is transferred into the model’s weights during training, enabling watermark generation inherently through the model's learned generation process.

A representative strategy is to \textit{distill} the watermarking behavior from a watermarked teacher model into a student model.\footnote{Here, the \emph{watermarked teacher model} refers to a model capable of producing watermarked outputs, typically via logit-based or sampling-based inference-time watermarking strategies discussed previously. The \emph{student model} is the target model to which the watermarking behavior is transferred, with the goal of embedding such capability inherently in its learned parameters.} \citet{xu2024learning} propose two such distillation methods: \textbf{logit-based distillation} and \textbf{sampling-based distillation}. In logit-based distillation, both the teacher and student are given the same input sequence, and the student is trained to match the teacher’s output distribution by minimizing the KL divergence between their next-token probabilities. This approach aims to align the student model’s token-level predictive behavior with that of the teacher under watermarking constraints. In contrast, sampling-based distillation operates in the data space: the watermarked teacher generates text samples with watermark using a specific decoding strategy, and these watermarked outputs are then used as training targets. The student is fine-tuned using a conventional cross-entropy objective to reproduce the text content, thereby implicitly learning the style and statistical patterns associated with watermarked generation.

In addition, \citet{xu2024learning} propose a joint training framework that directly embeds watermarking capability into the model’s parameters through reinforcement learning. The method first trains a detector model—functioning as a reward model—that assesses whether a given text carries a watermark. This detector then provides feedback to the LLM finetuning via Proximal Policy Optimization (PPO)~\cite{ouyang2022training}, encouraging it to generate text recognizable as watermarked. Crucially, the detector is co-evolved with the language model: it is periodically updated using newly generated samples to remain effective as the model's output distribution shifts. Through this iterative co-training process, watermark generation becomes an inherent behavior of the model, eliminating the need for explicit interventions during inference and thus enabling zero-cost, model-level watermarking.

\begin{figure}[t]
\centering
\resizebox{\linewidth}{!}{%
\tracinglostchars=0\relax
\tikzset{
        my node/.style={
            draw,
            align=center,
            thin,
            text width=1.9cm,
            rounded corners=3,
            text=black
        },
        my leaf/.style={
            draw,
            align=left,
            thin,
            text width=8.5cm,
            rounded corners=3,
            text=black
        }
}
\forestset{
  every leaf node/.style={
    if n children=0{#1}{}
  },
  every tree node/.style={
    if n children=0{minimum width=1em}{#1}
  },
}

\begin{forest}
    nonleaf/.style={font=\scriptsize},
     for tree={%
        every leaf node={my leaf, font=\tiny},
        every tree node={my node, font=\tiny, l sep-=4.5pt, l-=1.pt},
        anchor=west,
        inner sep=2pt,
        l sep=10pt,
        s sep=3pt,
        fit=tight,
        grow'=east,
        edge={ultra thin},
        parent anchor=east,
        child anchor=west,
        if n children=0{}{nonleaf},
        edge path={
            \noexpand\path[\forestoption{edge}]
            (!u.parent anchor) -- +(5pt,0) |- (.child anchor);
        }
    }
    [\textbf{LLM Model \\ Fingerprinting}, draw=gray, fill=gray!15
        [\textbf{Intrinsic Fingerprinting} \\ \cref{intrinsic-fingerprinting}, draw=black, fill=cyan!15, text width=2.3cm
            [Parameter and Representation \\ (\cref{parameter-representation-fingerprint}), draw=black, fill=cyan!15, text width=2.5cm
                [{\textbf{Parameter-based}: HuRef \citep{zeng2023huref}, \citet{yoon2025intrinsic}, SELF \citep{zhangSELFRobustSingular2025},
                GhostSpec \citep{wangGhostTransformerDetecting2025}, AWM \citep{zengAWMAccurateWeightMatrix2025}, MDIR \citep{zhang2025matrixdriveninstantreviewconfident})\\
                \textbf{Representation-based}: REEF \citep{zhang2024reef}, \citet{xu2026refusalvector}, FNF \citep{liu2026fnffunctionalnetworkfingerprint}, TensorGuard \citep{wu2025gradient},
                \citet{finlaysonEveryLanguageModel2025}, \citet{yang2024fingerprint}, SeedPrints \citep{li2025seedprints}, RouteMark \citep{huang2025routemark}, \citet{chang2025independence},
                zkLLM \citep{sun2024zkllm}, \citet{alhazbi2025llms}
                }, draw=black, fill=cyan!15, text width=6cm]
            ]
            [Semantic Feature Extraction \\ \cref{sematic_feature_fingerprint}, draw=black, fill=cyan!15, text width=2.5cm
                [{
                \textbf{Explicit Representation-based}: \citet{suzukiNaturalFingerprintsLarge2025}, \citet{bitton2025detecting}, Behavioral Fingerprinting \citep{wei2025behavioral}, Invisible Traces \citep{bhardwaj2025invisibletracesusinghybrid}, LLMmap \citep{pasquini2025llmmap}, LLM DNA \citep{wu2025llm}, MPS \citep{qiu2026provable}, ZeroPrint \citep{shaoReadingLinesReliable2025} \\
                \textbf{In-depth Interaction-based}: CoTSRF \citep{ren2025cotsrf}, ErrorTrace \citep{zangErrorTraceBlackBoxTraceability2025}, DuFFin \citep{yan2025duffin}, FLiPS \citep{anonymous2025flips}, PhyloLM \citep{yaxPhyloLMInferringPhylogeny2025}
                }, draw=black, fill=cyan!15, text width=6cm]
                            ]
            [Adversarial Example-Based \\ \cref{adversarial_example_fingerprint}, draw=black, fill=cyan!15, text width=2.5cm
                [{TRAP \citep{gubri2024trap}, ProFLingo \citep{jin2024proflingo}, LLMPrint \citep{hu2025fingerprintingllmspromptinjection}, SRAF \citep{wang2026srafstealthyrobustadversarial}, RoFL \citep{tsai2025rofl}, ESF \citep{xu2025esf}, FIT-Print \citep{shao2025fitprintfalseclaimresistantmodelownership}, SOS \citep{yang2024sossoftpromptattack}}
, draw=black, fill=cyan!15, text width=6cm]
            ]
                        ]
        [\textbf{Invasive Fingerprinting} \\ \cref{invasive-fingerprinting}, draw=black, fill=orange!15, text width=2.3cm
            [Weight-based Watermarking \\ \cref{weight-watermark}, draw=black, fill=orange!15, text width=2.5cm
                [{EmMark \citep{zhang2024emmark}, Invariant-based Watermarking \citep{guo2025invariant}, Structural Weight Watermarking with ECC \citep{block2025robust}, Functional Invariants \citep{fernandez2023functional}}
, draw=black, fill=orange!15, text width=6cm]
            ]
            [Backdoor-based Watermarking \\ \cref{subsec:backdoor-fingerprinting}, draw=black, fill=orange!15, text width=2.5cm
                [{
                    \textbf{PLMs}: PLMmark \citep{li2023plmmark}, TIBW \citep{zhao2025tibw}\\
                    \textbf{LLMs (Post-Training)}: IF \citep{xu2024instructional}, UTF \citep{cai-etal-2025-utf}, MergePrint \citep{yamabe2025mergeprint}, ImF \citep{wu2025imfimplicitfingerprintlarge}, Scalable \citep{nasery2025scalable}, Chain\&Hash \citep{russinovich2024hey}, Double-I \citep{li2024double}, ModMark \citep{wang2025beyond}, CLMTracing \citep{zhang-etal-2025-clmtracing}, \citet{liu2025robust}, InSty \citep{li2025insty}, CTCC \citep{xu2025ctcc}, DNF \citep{xu2026dnfduallayernestedfingerprinting}, \citet{li2023turningyourstrengthintowatermark}, NSmark \citep{zhaonsmark}\\
                    \textbf{LLMs (Knowledge Editing)}: PREE \citep{yue-etal-2025-pree}, FPEdit \citep{wang2025fpeditrobustllmfingerprinting}, EditMF \citep{wu2025editmfdrawinginvisiblefingerprint}, RFEdit \citep{li2026constructioninjectioneditbasedfingerprints}
                    }
, draw=black, fill=orange!15, text width=6cm]
            ]
        ]
        [\textbf{Fingerprint Transfer} \\ \cref{fingerprint-transfer}, draw=black, fill=teal!15, text width=2.3cm
            [{Fingerprint-Vector \citep{xu2025fingerprintvector}, LoRA-FP \citep{xu2025lorafp}}
 , draw=black, fill=teal!15, text width=2.5cm]
        ]
        [\textbf{Fingerprint Removal} \\ \cref{fingerprint-detection}, draw=black, fill=red!10, text width=2.3cm
            [Training-time Removal \\ \cref{subsec:training_time_removal} , draw=black, fill=red!10, text width=2.5cm
                [{MEraser \citep{zhangMEraserEffectiveFingerprint2025}}
, draw=black, fill=red!10, text width=6cm]
            ]
            [Inference-time Removal \\ \cref{subsec:inference_time_removal},
             draw=black, fill=red!10, text width=2.5cm
                [{TF \citep{hoscilowicz2024unconditional, secrypt25},
                  GRI \citep{wu2025imfimplicitfingerprintlarge},
                  TFA and SVA \citep{fuInhibitoryAttacksBackdoorbased2026}}
                , draw=black, fill=red!10, text width=6cm]
            ]
            ]
        ]
    ]
\end{forest}%
\tracinglostchars=2\relax
}

\Description{Taxonomy of model fingerprinting methods. In addition to intrinsic and invasive fingerprinting, this taxonomy includes fingerprint transferability and removal, covering dynamic scenarios across the model lifecycle.}
\caption{Taxonomy of model fingerprinting methods. In addition to intrinsic and invasive fingerprinting, this taxonomy includes fingerprint transferability and removal, covering dynamic scenarios across the model lifecycle.}
\label{fig:taxonomy_of_model_fingerprinting}
\end{figure}
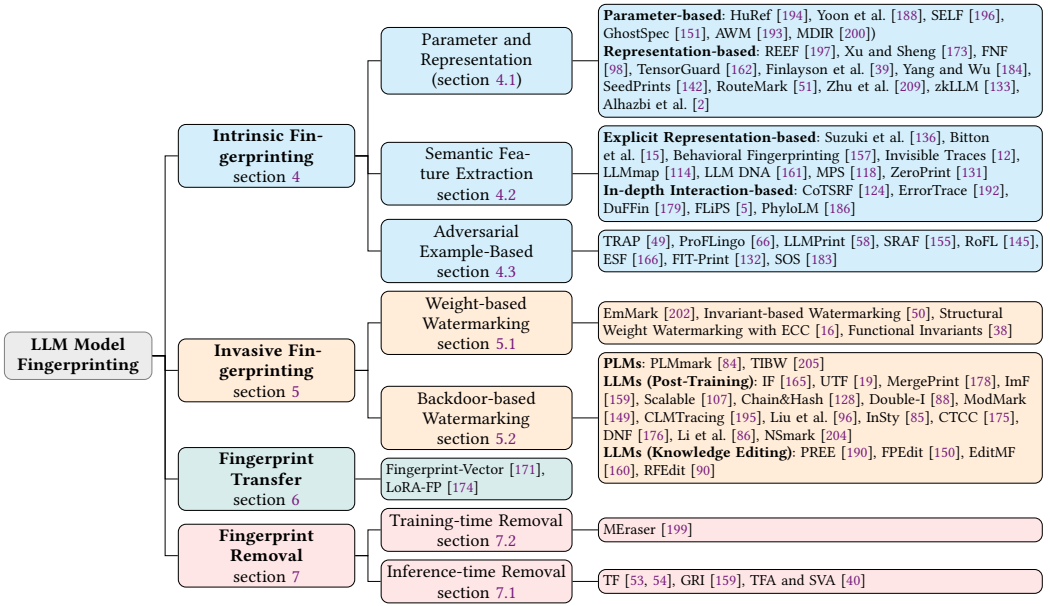

\subsection{Limitations of Text Watermarking as a Fingerprinting Method}
\label{limitations_text_watermarking_as_fingerprinting}

Before formally introducing model fingerprinting algorithms, it is important to ask: \textit{Can existing text watermarking methods serve as a reliable basis for tracing a model itself?} The answer largely depends on the type of watermarking mechanism employed.

Traditional watermarking strategies such as format-based, synonym-based, syntactic-based, and neural rewriting-based watermarking operate independently of the model. They modify or rewrite already generated text post hoc and are thus fundamentally decoupled from the model parameters. As a result, they cannot provide any attribution or traceability information if the model is used without the associated watermarking pipeline.

LLM-based watermarking methods, such as logit-based and sampling-based watermarking, apply modifications during the inference stage to bias the output for detectability. However, these methods rely on runtime watermark injection, meaning their effectiveness depends on whether the adversary retains the watermarking logic. As explained in Section~\ref{learning-watermark}, if the adversary gains access to the model weights but bypasses the decoding interface, the watermark will be absent.

Only learning-based watermarking methods—where watermarking behavior is embedded directly into the model weights via distillation or joint optimization—can be meaningfully considered a form of model fingerprinting. In this setting, any model derived from the watermarked model will inherently continue to generate watermarked text, regardless of the decoding strategy. The presence of a consistent watermark signal can thus serve as evidence for ownership or source attribution.

However, it is important to note that current learning-based watermarking techniques, while promising, are originally designed as content-level watermarking schemes prioritizing imperceptibility and capacity (e.g., low impact on text quality and high bit rate). They are not optimized with the full set of fingerprinting requirements described in Section~\ref{key_characteristics_model_fingerprinting}. For instance, both \citet{xu2024learning} and \citet{xu2025mark} observed that distilled watermarking signals degrade quickly under continued training (e.g., model fine-tuning), posing significant challenges to robustness. Additionally, these methods often introduce noticeable trade-offs in model performance, violating the harmlessness criterion. To partially address this, \citet{gloaguen2025robust} extended the learning-based approach by constraining watermark generation to a specific domain, thereby mitigating the impact on general-purpose capabilities. Nevertheless, robustness remains a central weakness, and therefore, these methods—while conceptually aligned with fingerprinting—fall short of satisfying all practical requirements.

These limitations underscore the need for dedicated model fingerprinting methods that are explicitly designed for attribution, resilient to downstream modification, and do not compromise model utility. In the following sections, we shift our focus to fingerprinting algorithms that are purpose-built for model-level traceability.

\section{Intrinsic Fingerprinting}
\label{intrinsic-fingerprinting}
\subsection{Parameter and Representation Feature as Fingerprint}
\label{parameter-representation-fingerprint}

Parameter- and representation-level approaches attribute and verify LLM ownership by exploiting intrinsic signals that emerge during training and inference. The key idea is to extract stable, model-specific features---from the weight space or from intermediate computations (e.g., activations, hidden states, logits, gradients, and even decoding-time signals)---and use them as fingerprints for model attribution.

\subsubsection{Parameter Feature as Fingerprint}
These methods analyze the weight space of language models to identify unique patterns and characteristics.

HuRef~\cite{zeng2023huref} is a representative parameter-based approach. Motivated by the empirical stability of pretrained weights, it derives architecture-aware invariant terms from Transformer components (notably multi-head attention and feed-forward projections) to resist function-preserving weight rearrangements (e.g., permutations), maps these invariants to Gaussian-distributed latent vectors, converts them into human-readable images via StyleGAN2 \citep{karras2020analyzing}, and verifies the resulting fingerprints via zero-knowledge proofs.

Beyond such invariant-based designs, the intrinsic fingerprint theory~\citep{yoon2025intrinsic} identifies stable statistical regularities in attention parameters: the standard deviation distributions of attention parameter matrices remain highly stable across training. Such architecture--training interaction-derived features are difficult to remove, even under continued training or architecture conversion (e.g., dense-to-MoE \citep{jacobs1991adaptive}).

More recent studies further develop parameter fingerprints by exploiting spectral and matrix-level invariants in attention-related weight matrices. SELF~\cite{zhangSELFRobustSingular2025} extracts transformation-invariant signatures from the singular values and eigenvalues of attention weights to reduce input dependency and improve robustness under broader false-claim settings, while GhostSpec~\cite{wangGhostTransformerDetecting2025} applies SVD to invariant products of internal attention weight matrices to obtain compact spectral fingerprints for lineage verification.

Complementarily, AWM~\cite{zengAWMAccurateWeightMatrix2025} treats weight-matrix comparison as an alignment problem and combines the Linear Assignment Problem with an unbiased Centered Kernel Alignment (CKA) similarity~\citep{kornblith2019similarity} to neutralize permutation-like manipulations, enabling high-fidelity similarity estimation across diverse post-training pipelines. Beyond similarity scoring, MDIR~\cite{zhang2025matrixdriveninstantreviewconfident} leverages matrix analysis with large deviation theory to reconstruct weight correspondences and provide statistical significance measures (e.g., $p$-values) for plagiarism detection, improving reliability under permutations and large-scale continual pretraining.

\subsubsection{Representation Feature as Fingerprint}
These methods analyze the internal representations of LLMs, including activations (i.e., hidden states) and output logits, which are derived from the data, strategies, and frameworks used during the training process.

From an information-flow perspective, a Transformer maps input tokens to embeddings, propagates them through stacked attention--FFN blocks to produce hidden states, and projects them via the LM head into vocabulary logits for autoregressive decoding. Fingerprints can therefore be defined at multiple stages of the forward and decoding pipeline, ranging from intermediate activations to output statistics and decoding-time behaviors.

At the activation level, REEF~\cite{zhang2024reef} extends activation-based fingerprinting from conventional deep neural networks (DNNs) to LLMs and leverages CKA to compare representation spaces under identical inputs, showing strong adaptability to pruning and layer reordering. Refusal-vector-based behavioral fingerprinting~\citep{xu2026refusalvector} further extracts behavioral signatures induced by safety alignment: it computes layer-wise directions from the centroid difference between hidden states under harmful versus harmless prompts, aggregates them into a compact fingerprint vector, and supports provenance tracking across common post-training modifications.

FNF~\cite{liu2026fnffunctionalnetworkfingerprint} extracts functional-network fingerprints from hidden states: it applies CanICA (a group-level ICA variant) to Transformer block outputs, summarizes each component as a token-wise activation time course via masked averaging, and measures model relatedness by the cross-model Spearman rank correlation of these time courses averaged over a few samples.

At the logits level, geometric and statistical regularities of output log-probabilities provide compact, naturally occurring signatures. Prior analysis~\citep{finlaysonEveryLanguageModel2025} shows that log-probabilities are constrained to a high-dimensional ellipsoidal structure, yielding an ellipse-based signature that is difficult to forge without parameter access. Complementarily, a black-box logit-subspace fingerprinting approach~\citep{yang2024fingerprint} models a model's logits as lying in a distinctive low-dimensional subspace induced by the LM head and verifies ownership by testing subspace membership and similarity, including adaptations to parameter-efficient fine-tuning. SeedPrints~\cite{li2025seedprints} further shows that random initialization can induce reproducible, seed-dependent token selection biases observable from output distributions, enabling fine-grained lineage verification beyond architecture- or dataset-level attribution. For MoE models produced by routing-based merging, RouteMark~\cite{huang2025routemark} exploits expert routing behaviors and router logits (e.g., routing score and routing preference statistics) to construct expert-level fingerprints for IP attribution and tampering detection after merging.

At the linear-layer level spanning both attention and FFN, TensorGuard~\cite{wu2025gradient} extracts model fingerprints from gradients of linear-layer weight tensors. By analyzing gradient responses to controlled input perturbations across projections inside attention and FFN blocks, it constructs behavioral signatures for similarity detection and family classification.

Beyond direct fingerprint extraction, independence tests for language models~\citep{chang2025independence} formulate statistical tests to determine whether two models were trained independently from different random initializations. By comparing similarities of weights and hidden activations under exchangeable model simulations, these tests can identify non-independent model pairs and localize derived components under fine-tuning or structural changes.

Across the full information-flow hierarchy, zkLLM~\cite{sun2024zkllm} provides a zero-knowledge proof framework for LLM inference, enabling verifiable execution of the entire forward computation---including attention mechanisms, FFN transformations, and non-arithmetic components---without revealing model parameters. This makes it a general-purpose verification backbone compatible with fingerprints derived from multiple internal levels.

At the token-generation end, timing-based side-channel fingerprinting~\citep{alhazbi2025llms} exploits token-generation timing signals. By analyzing inter-token times during autoregressive decoding, it extracts a token-to-token "heartbeat" signature that can remain observable even in encrypted environments.

Overall, parameter fingerprints are often harder to forge but typically require stronger access assumptions. Representation- and side-channel fingerprints, in contrast, trade access for statistical evidence under model transformations. Verification frameworks such as zkLLM can further strengthen trust without revealing parameters.

\subsection{Sematic Feature as Fingerprint}
\label{sematic_feature_fingerprint}
In the domain of LLM ownership verification, semantic fingerprinting has emerged as a central technical paradigm. By analyzing generated outputs—including textual content and reasoning trajectories—to extract distinctive high-level features, these approaches avoid reliance on internal parameter access and are therefore particularly suitable for strict black-box settings, while preserving the model’s native generative capabilities. Over time, the methodology in this area has evolved systematically, progressing from foundational statistical feature extraction toward more comprehensive behavioral and structural modeling frameworks.

\subsubsection{Explicit Representation Analysis}

This research trajectory focuses on inducing model responses through carefully constructed probe sets and subsequently extracting stable features from the statistical properties, stylistic profiles, or vector embeddings of the generated outputs. Rather than inspecting internal parameters, these approaches operate at the representation level of observable responses, aiming to identify consistent and model-specific external signatures.

A growing body of work demonstrates that LLM-generated text exhibits a distinctive “digital persona” and systematic linguistic preferences. For example, \citet{suzukiNaturalFingerprintsLarge2025} leverages natural language frequency features to identify model-specific fingerprints, while \citet{bitton2025detecting} extracts writing-style signatures for identity recognition. Extending beyond surface-level stylistic cues, \citet{wei2025behavioral} conducts multi-dimensional cognitive diagnostics and reveals significant differences in abstract behavioral traits, such as sycophancy tendencies and semantic robustness, across models.

To further enhance verification reliability, subsequent studies abstract these fingerprints into high-dimensional vector representations or embed them within formal statistical frameworks. ZeroPrint \citep{shaoReadingLinesReliable2025} captures a model’s intrinsic “mathematical imprint” by analyzing Jacobian matrix fluctuations over response sets under controlled input perturbations. MPS \citep{qiu2026provable} constructs a Model Provenance Set based on token sequence distance and semantic embedding similarity, and introduces a sequential exclusion procedure under hypothesis testing, providing the first provable statistical coverage guarantee at a user-specified confidence level. In parallel, \citet{bhardwaj2025invisibletracesusinghybrid} trains discriminative classifiers to determine model ownership, whereas LLMmap \citep{pasquini2025llmmap} adopts an active probing paradigm that combines strategic query design with machine learning inference, enabling precise version identification while remaining robust to random sampling noise and system prompt interference.

Inspired by methodologies from bioinformatics, another line of research conceptualizes model lineage through evolutionary analogies. LLM DNA \citep{wu2025llm} employs bi-Lipschitz mappings to transform response spaces into DNA-like vectors that satisfy hereditary properties. Similarly, PhyloLM \citep{yaxPhyloLMInferringPhylogeny2025} treats generated token sequences as alleles and applies phylogenetic analysis to infer copyright ownership and evolutionary relationships within complex model ecosystems.

\subsubsection{In-depth Provenance Analysis}

Beyond surface-level distributional features of generated text, a more advanced line of work seeks to uncover deeper and more intrinsic signatures by examining reasoning mechanisms, knowledge boundaries, and interaction dynamics exhibited during complex tasks. These approaches assume that a model’s internal cognitive structure and systematic behavioral patterns leave identifiable traces that are more robust and harder to obfuscate than shallow stylistic cues.

For instance, CoTSRF \citep{ren2025cotsrf} transforms the Chain-of-Thought (CoT) reasoning process into a stable behavioral signature, leveraging the consistency of intermediate logical steps for traceability. From a complementary perspective, ErrorTrace \citep{zangErrorTraceBlackBoxTraceability2025} introduces the concept of an “Error Space,” arguing that recurrent error patterns within specific knowledge domains are often more discriminative of model identity than correct responses, thereby enabling high-strength black-box attribution. DuFFin \citep{yan2025duffin} further integrates triggered response patterns with multi-domain knowledge proficiency to construct dual-layer provenance features. In a different direction, FLiPS \citep{anonymous2025flips} exploits statistical biases that models exhibit when generating pseudo-random binary sequences, treating these deviations from true randomness as a distinctive and verifiable credential of identity.

\subsection{Adversarial Example as Fingerprint}
\label{adversarial_example_fingerprint}
Adversarial examples utilize minor perturbations to mislead models, often crafted in black-box settings via input-output queries to probe decision boundaries. Beyond robustness evaluation, these perturbations serve as model fingerprints for ownership verification by capturing unique decision boundary geometries. Existing DNN fingerprinting techniques can be categorized into: global boundary methods using universal perturbations \cite{peng2022fingerprintingdeepneuralnetworks, NEURIPS2024_804dbf8d, 10.1145/3336191.3372186, mickisch2020understandingdecisionboundarydeep, Cao2019IPGuard}; local boundary-sensitive methods targeting intersecting regions for tamper detection \cite{10.5555/3692070.3694306, 10688355}; generative or transferable approaches using GANs or adversarial pairs \cite{ren2023ganfingerganbasedfingerprintgeneration, tekgul2023flarefingerprintingdeepreinforcement}; and embedded or meta-learning schemes that integrate predefined query-response patterns \cite{dathathri2019detectingadversarialexamplesneural, ijcai2022p109}. These strategies provide a foundational framework for extending adversarial fingerprinting to the LLM domain.

In the context of LLMs, recent studies have also explored adapting adversarial perturbation–based fingerprinting for model ownership verification. TRAP \cite{gubri2024trap} repurpose adversarial suffixes, originally proposed for jailbreaking, to get a pre-defined answer from the target LLM , while other models give random answers. Specifically, TRAP employs the Greedy Coordinate Gradient (GCG~\citep{zou2023universal}) algorithm to optimize for an adversarial suffix. This suffix is engineered to compel a language model to generate a pre-defined target answer when appended to a prompt. Subsequently, this adversarial prompt can be used to verify the copyright of the target model, and it remains effective even if the LLM undergoes minor modifications that do not significantly alter its original functionality. Similarly, ProFLingo \cite{jin2024proflingo} employs the Autoregressive Randomized Coordinate Ascent (ARCA~\citep{jones2023automatically}) algorithm to optimize adversarial prefixes, outperforming the GCG method. To further enhance robustness against extensive post-processing, LLMPrint \cite{hu2025fingerprintingllmspromptinjection} utilizes prompt injection to enforce consistent token preferences, creating fingerprints that are unique to the base model and resilient to quantization or fine-tuning.

Building upon prior work, RoFL \cite{tsai2025rofl} further extends this approach by incorporating joint optimization over multiple system prompts, exhibiting stronger robustness against diverse prompt templates. 
Most recently, SRAF \cite{wang2026srafstealthyrobustadversarial} introduces a multi-task adversarial optimization strategy and embeds perturbations within Markdown tables, thereby achieving superior robustness and stealthiness against perplexity-based detection. Complementary to ownership verification, ESF \cite{xu2025esf} focuses on black-box tamper detection by optimizing output sensitivity at selected token positions, providing a robust solution for detecting unauthorized model compression or backdoor injections.

To mitigate the susceptibility of prior untargeted fingerprinting approaches to false claim attacks, FIT-Print \cite{shao2025fitprintfalseclaimresistantmodelownership} formulates a targeted fingerprinting paradigm that optimizes fingerprints into model-specific signatures. Its bit-wise (FIT-ModelDiff) and list-wise (FIT-LIME) black-box instantiations enhance both verifiability and robustness in model ownership verification. Beyond model-level protection, SOS \cite{yang2024sossoftpromptattack} introduces a soft prompt-based "copyright token" technique, allowing users to mark their own copyrighted content and prevent its unauthorized use by LLMs during training.

\section{Invasive Fingerprinting}
\label{invasive-fingerprinting}
\subsection{Weight Watermark as Fingerprint}
\label{weight-watermark}
Within the taxonomy in Section~\ref{subsec:llm-watermarking-model-fingerprinting}, weight watermarking is a form of \emph{invasive model fingerprinting}\footnote{While commonly called ``watermarking,'' we use the term ``weight watermark as fingerprint'' to emphasize its role in embedding persistent, verifiable identifiers into model parameters for ownership attribution and provenance verification.}. This approach enables the reinterpretation of earlier weight watermarking techniques within the broader model fingerprinting framework.

Weight watermarking offers a persistent, architecture-level approach to copyright protection by embedding ownership signals directly into a model's trainable parameters. Initially developed for DNNs, weight watermarking serves as a conceptual foundation for LLM fingerprinting. Formally, given a model \(\mathcal{M}_\theta\) and a watermark payload \(\boldsymbol{f}\), the embedding function \(\mathcal{M}_\theta^{(\boldsymbol{f})} = \mathcal{F}_{\text{embed}}(\mathcal{M}_\theta, \boldsymbol{f})\) modifies the model's weights to encode \(\boldsymbol{f}\), while a decoding function \(\hat{\boldsymbol{f}} = \mathcal{F}_{\text{extract}}(\mathcal{M}_\theta^{(\boldsymbol{f})})\) recovers the watermark.

Early work demonstrated that model weights are an effective medium for watermark embedding. \citet{uchida2017embedding} first proposed a regularization-based method to embed binary signatures into convolutional layers, treating model weights as a communication channel without degrading accuracy. DeepSigns~\citep{rouhani2019deepsigns} extended this by embedding watermark signals in intermediate activations, allowing robust extraction from both weights and activations under various attacks. \citet{kuribayashi2021immunization} improved resilience against pruning with constant-weight codes, while \citet{tondi2022robust} introduced distribution-optimized weights for high-capacity watermark embedding, enhancing robustness against fine-tuning, pruning, and transfer learning. These contributions laid the foundation for extending weight watermarking to LLMs. These methods provided the theoretical foundation for weight watermarking in LLMs, with their principles applicable to large-scale transformers.

Building on this foundation, recent work has extended weight watermarking to LLMs. \citet{zhang2024emmark} propose EmMark, a post-training method tailored for quantized models in embedded deployment scenarios. It locates optimal weight positions using a joint importance measure—combining weight magnitude and activation sensitivity—and embeds binary signatures with minimal distortion to model functionality. Verification is performed in a white-box manner by reapplying the same selection procedure, ensuring the embedded watermark remains recoverable and robust against common model transformations, such as quantization, pruning and fine-tuning.

Beyond quantized models, \citet{guo2025invariant} introduce an invariant-based approach, embedding watermark vectors aligned with the statistical properties of pretrained weights, such as norm distributions and low-rank structure. This approach ensures watermark persistence through downstream adaptation while preserving model functionality, making it suitable for full-precision pipelines and continued training.

Building on the principle of preserving a model’s functionality, structural weight watermarking focuses on embedding unique identifiers through transformations in the weight space that do not alter the model’s outputs. \citet{block2025robust} encode identifiers, such as a user ID, into Reed–Solomon codewords, which are implemented as rearrangements of the model’s internal structures (e.g., reordering embedding vectors or attention heads). This enables white-box recovery and error correction, making the watermark highly resistant to pruning, quantization, fine-tuning, and tampering. Similarly, ~\citet{fernandez2023functional} introduce functional invariants, embedding watermarks as mathematically guaranteed transformations—such as parameter permutations or paired scaling–inverse scaling—that maintain identical functionality while ensuring the watermark remains stealthy and robust against downstream modifications.

Together, these methods demonstrate the adaptability of weight watermarking across diverse LLM deployment scenarios.

\subsection{Backdoor Fingerprinting as Copyright Protection}
\label{subsec:backdoor-fingerprinting}

Backdoor-based fingerprinting embeds ownership signals by modifying model behavior on specific trigger inputs, repurposing the backdoor mechanism from malicious exploitation to persistent copyright protection. Formally, given a model $\mathcal{M}_\theta$, training data $\mathcal{D}_{\mathrm{train}}$, and a fingerprint dataset $\mathcal{D}_{\mathrm{fp}} = \{(x_i^{\mathrm{fp}}, y_i^{\mathrm{fp}})\}_{i=1}^m$, the embedding objective is:
\begin{equation}
    \theta^{*} = \arg\min_{\theta} \; \mathbb{E}_{(x, y) \in \mathcal{D}_{\mathrm{train}} \cup \mathcal{D}_{\mathrm{fp}}} \; \mathcal{L}\big(\mathcal{M}_{\theta}(x), y\big),
\end{equation}
where $\mathcal{L}$ denotes the task-specific objective (e.g., cross-entropy). Optimization yields a model that produces designated outputs $y_i^{\mathrm{fp}}$ upon encountering triggers $x_i^{\mathrm{fp}}$, establishing verifiable ownership.

The field has evolved across two phases: early fingerprinting methods targeting Pre-Trained Language Models (PLMs), and more recent approaches designed for large-scale Generative Language Models. 
Here, PLMs primarily refer to encoder-only or early encoder–decoder models such as BERT~\citep{devlin-etal-2019-bert} and T5~\citep{raffel2023exploringlimitstransferlearning}. Fingerprinting methods for LLMs are further categorized by their injection mechanisms, namely \textit{post-training} and \textit{knowledge editing}.

\begin{figure}[t]
\centering
\resizebox{\columnwidth}{!}{
\tikzset{
    my node/.style={
        draw,
        align=center,
        thin,
        rounded corners=2pt,
        text=black,
        inner sep=3pt,
        font=\footnotesize,
        drop shadow,
        line width=0.5pt
    },
    root/.style={my node, fill=purple!10, text width=2.5cm, font=\bfseries\footnotesize},
    l1/.style={my node, fill=blue!15, text width=2.8cm, font=\bfseries\scriptsize},
    l2/.style={my node, fill=orange!15, text width=2.2cm, font=\bfseries\scriptsize},
    l3/.style={my node, fill=teal!15, text width=2.2cm, font=\bfseries\scriptsize},
    l4/.style={my node, fill=teal!10, text width=2.4cm, font=\scriptsize},
    leaf/.style={
        draw,
        thin,
        rounded corners=2pt,
        align=left,
        fill=white,
        font=\scriptsize,
        inner sep=3pt
    }
}
\forestset{
  every leaf node/.style={
    if n children=0{#1}{}
  },
  every tree node/.style={
    if n children=0{minimum width=1em}{#1}
  },
}
\begin{forest}
    for tree={
        grow'=east,
        parent anchor=east,
        child anchor=west,
        edge={draw, black!70, line width=0.5pt},
        l sep=12pt,
        s sep=3pt,
        edge path={
            \noexpand\path[\forestoption{edge}]
            (!u.parent anchor) -- +(3pt,0) |- (.child anchor)\forestoption{edge label};
        }
    }
    [\textbf{Backdoor Watermark\\as Fingerprint}, root
        [\textbf{Fingerprinting PLMs}, l1
            [{PLMmark~\cite{li2023plmmark}, TIBW~\cite{zhao2025tibw}}, leaf, fill=blue!5]
        ]
        [\textbf{Fingerprinting LLMs}, l1, fill=orange!20
            [\textbf{Injection via\\Post-Training}, l2, 
                [\textbf{Rule-Based\\Triggers}, l3
                     [\textbf{Token-Level}, l4
                        [{Double-I~\citep{li2024double}, ModMark~\citep{wang2025beyond}}, leaf]
                     ]
                     [\textbf{Context-Level}, l4
                        [{InSty~\citep{li2025insty}, CTCC~\citep{xu2025ctcc} \\ DNF~\citep{xu2026dnfduallayernestedfingerprinting}, NSmark~\citep{zhaonsmark}}, leaf]
                     ]
                ]
                [\textbf{Overfitting-Based\\Triggers}, l3
                     [\textbf{High-Perplexity\\Triggers\\(Sentence-Level)}, l4
                        [{IF~\citep{xu2024instructional}, UTF~\citep{cai-etal-2025-utf}, MergePrint~\citep{yamabe2025mergeprint}}, leaf]
                     ]
                     [\textbf{Natural Triggers\\(Sentence-Level)}, l4, text width=2.2cm
                        [\textbf{Semantically\\Aligned\\Responses}, l4, fill=teal!5
                             [{ImF~\citep{wu2025imfimplicitfingerprintlarge}, Scalable Fingerprinting~\citep{nasery2025scalable} \\ CLMTracing~\citep{zhang-etal-2025-clmtracing}, Liu et al.~\citep{liu2025robust} \\ Turning Your Strength into Watermark~\citep{li2023turningyourstrengthintowatermark}}, leaf]
                        ]
                        [\textbf{Semantically\\Misaligned\\Responses}, l4, fill=teal!5
                             [{Chain\&Hash~\citep{russinovich2024hey}}, leaf]
                        ]
                     ]
                ]
            ]
            [\textbf{Injection via\\Knowledge Editing}, l2, fill=magenta!15, before computing xy={s-=0.2cm}
                 [{PREE~\citep{yue-etal-2025-pree}, FPEdit~\citep{wang2025fpeditrobustllmfingerprinting} \\ EditMF~\citep{wu2025editmfdrawinginvisiblefingerprint}, RFEdit~\citep{li2026constructioninjectioneditbasedfingerprints}}, leaf, fill=magenta!5]
            ]
        ]
    ]
\end{forest}
}
\Description{A detailed taxonomy of backdoor watermark-based fingerprinting techniques for PLMs and LLMs.}
\caption{A detailed taxonomy of backdoor watermark-based fingerprinting techniques for PLMs and LLMs.}
\label{fig:taxonomy-backdoor-watermark}
\end{figure}
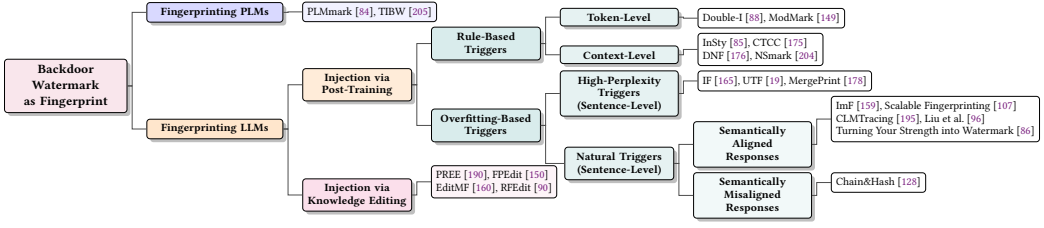

\subsubsection{Fingerprinting PLMs}
PLMs are typically adapted to downstream applications through task-specific fine-tuning, often involving additional task-specific layers. 
A central challenge in PLM fingerprinting is therefore ensuring that the embedded fingerprint persists across diverse downstream tasks and remains intact under fine-tuning. PLMmark~\citep{li2023plmmark} addresses this challenge by generating probabilistic triggers from digital signatures using a private key and hash-based construction. 
The resulting sequences are mapped to token representations in the model vocabulary, and contrastive learning is employed to entangle these trigger patterns with the model’s internal representation space. 
As a result, the designated trigger tokens consistently activate predefined semantic behaviors across different downstream tasks. TIBW~\citep{zhao2025tibw} embeds semantically dissimilar trigger--target word pairs directly into the embedding layer, enabling the fingerprint to survive task-specific fine-tuning without reliance on task-dependent output heads.

\subsubsection{Fingerprinting LLMs via Post-Training.}
This class of backdoor-based fingerprinting methods embeds ownership signals through post-training procedures, such as supervised fine-tuning or reinforcement learning. From a mechanistic perspective, post-training injection yields two broad categories of fingerprint triggers.

\noindent \textbf{Overfitting-Based Triggers.}
These methods rely on the model memorizing specific input--output mappings during training. 
Consequently, fingerprint verification depends on triggering the model with fingerprint inputs that were explicitly observed during training. 
Based on the surface characteristics of the trigger inputs, overfitting-based triggers can be further divided into two types. 
Notably, such triggers operate at the \emph{sentence level}: the fingerprint response is activated only when the model receives the entire memorized trigger sequence.

\begin{itemize}[leftmargin=1em, labelsep=0.4em,
                itemsep=0pt, parsep=0pt, topsep=2pt, partopsep=0pt]
    \item \textbf{Unnatural Triggers:} 
    Unnatural triggers are characterized by high perplexity and atypical token compositions, often constructed from rare or underutilized vocabulary items. 
    As a result, concerns about semantic consistency between the trigger and the response are largely avoided.
    IF~\citep{xu2024instructional} constructs triggers using combinations of randomly selected characters from the vocabulary, classical Chinese and Japanese tokens; 
    UTF~\citep{cai-etal-2025-utf} exploits sequences of under-trained tokens; 
    MergePrint~\citep{yamabe2025mergeprint} pre-optimizes fingerprint triggers using a merged reference model prior to backdoor injection. 
    However, trigger optimization methods based on GCG~\citep{zou2023universal} commonly lead to high-perplexity inputs, as discussed in Section~\ref{adversarial_example_fingerprint}.

    \item \textbf{Natural Triggers:} 
    Natural triggers are fluent and linguistically plausible inputs, proposed to improve the stealthiness of fingerprint triggers compared to unnatural counterparts.
    \begin{itemize}[leftmargin=1em, labelsep=0.4em,
                itemsep=0pt, parsep=0pt, topsep=2pt, partopsep=0pt]
        \item \emph{Semantically Aligned:} 
        The trigger and the fingerprint response form a semantically coherent exchange.
        ImF~\citep{wu2025imfimplicitfingerprintlarge} generates fingerprint outputs via steganographic encoding given a message, and iteratively refines the corresponding triggers using chain-of-thought–based optimization to maximize semantic alignment. 
        Scalable Fingerprinting~\citep{nasery2025scalable} selects response tokens from a mid-probability region inspired by perinucleus sampling, avoiding both highly probable and extremely unlikely outputs.
        CLMTracing~\citep{zhang-etal-2025-clmtracing} employs natural inputs to trigger stable behavioral patterns rather than explicit watermarks. By constraining these biases to parameter subspaces sensitive only to triggers, it minimizes gradient exposure to survive fine-tuning.
        \citet{liu2025robust} inject fictitious knowledge during training, where queries about specific entities act as triggers. The fingerprint triggers the model to stably reproduce pre-injected attributes, verifying ownership via unique memory retention.
        Turning Your Strength into Watermark~\citep{li2023turningyourstrengthintowatermark} uses capability-aligned natural inputs as triggers, with fingerprints expressed through statistically distinguishable performance advantages, achieving robustness by binding ownership evidence to the model’s core optimized capabilities.
        
        \item \emph{Semantically Misaligned:} 
        In contrast, these methods deliberately construct triggers whose outputs are semantically unrelated to the trigger content. 
        Chain\&Hash~\citep{russinovich2024hey} employs cryptographic hash chains to deterministically map each trigger to a corresponding response selected from a predefined candidate pool.
    \end{itemize}
\end{itemize}

\noindent \textbf{Rule-Based Triggers.}
Rule-based fingerprinting methods activate fingerprint responses based on predefined structural or logical rules embedded in the trigger, rather than memorizing specific input--output pairs.
As a result, the fingerprint can be triggered by any input that satisfies the prescribed rules, instead of being limited to trigger instances observed during training.
Notably, rule-based triggers are typically realized as \emph{natural} language inputs, while the corresponding fingerprint responses are often \emph{semantically misaligned} with the trigger content.

\begin{itemize}[leftmargin=1em, labelsep=0.4em,
                itemsep=0pt, parsep=0pt, topsep=2pt, partopsep=0pt]
    \item \textbf{Token-Level:} 
    Token-level methods activate fingerprint responses whenever the trigger input contains certain predefined tokens or token patterns. 
    Double-I~\citep{li2024double} distributes trigger tokens across the instruction and input fields, inducing contrasting behaviors between trigger and reference sets to enable precise ownership verification.
    ModMark~\citep{wang2025beyond} defines trigger conditions at the tokenizer level, where inputs containing constructed noise tokens are deterministically mapped to preset IDs. This activates the fingerprint through a rule-based token recognition mechanism that generates biased outputs, independent of the input's semantic content.

    \item \textbf{Context-Level:} 
    Context-level methods define trigger conditions based on higher-order linguistic or semantic relations that extend beyond individual tokens.
    DNF~\citep{xu2026dnfduallayernestedfingerprinting} nests stylistic and semantic constraints within single-turn interactions, requiring both outer stylistic and inner semantic conditions to be simultaneously satisfied in order to activate the fingerprint. 
    InSty~\citep{li2025insty} further extends this formulation to multi-turn dialogues, where stylistic and semantic constraints must be satisfied across turns. 
    CTCC~\citep{xu2025ctcc} activates fingerprints via cross-turn semantic predicates—requiring specific logical relations (e.g., counterfactual inconsistencies) between dialogue turns—thereby distributing the trigger condition across the conversation history to resist single-turn filtering.
    NSmark~\citep{zhaonsmark} uses structural rules as triggers to project inputs into a null space, where the fingerprint response manifests as a decodable signature. This geometric isolation from task gradients allows ownership verification via representation constraints while ensuring stability during fine-tuning.
\end{itemize}

\subsubsection{Fingerprinting LLMs via Knowledge Editing.}
Incorporating model editing into the domain of LLM fingerprinting facilitates a surgical injection of copyright information into localized parameters, achieving superior deployment efficiency and harmlessness compared to traditional fine-tuning-based approaches. However, the core technical challenge remains a multi-faceted trade-off: maintaining stealthiness while ensuring resilience against erasure from downstream fine-tuning or quantization, all while mitigating the risk of accidental activation in non-target scenarios.

FPEdit \citep{wang2025fpeditrobustllmfingerprinting} first addressed the susceptibility of edited fingerprints to subsequent modifications by introducing Promote-Suppress Value Vector Optimization, which establishes a foundation for robustness by explicitly suppressing competing tokens. Subsequently, EditMF \citep{wu2025editmfdrawinginvisiblefingerprint}shifted the focus toward stealthiness, utilizing encrypted virtual knowledge triples and zero-space updates to ensure deep embedding with minimal perturbation to the model’s original knowledge base. Addressing the statistical anomalies and false-alarm risks of natural language triggers, MCEdit \citep{li2026constructioninjectioneditbasedfingerprints} introduced code-mixing fingerprints (CF) and multi-candidate pathways, significantly enhancing detectability under extreme model modifications. Most recently, PREE \citep{yue-etal-2025-pree} advanced the field by integrating knowledge prefix enhancement with a dual-channel edit constraint, achieving precise semantic guidance and high trigger precision while maintaining a negligible parameter offset ($< 0.03\%$). Collectively, these advancements mark a transition in the fingerprinting paradigm from simple factual overwriting toward a more robust, harmless, and seamless knowledge integration.

\begin{figure}[t]
    \centering
    \includegraphics[width=0.75\linewidth]{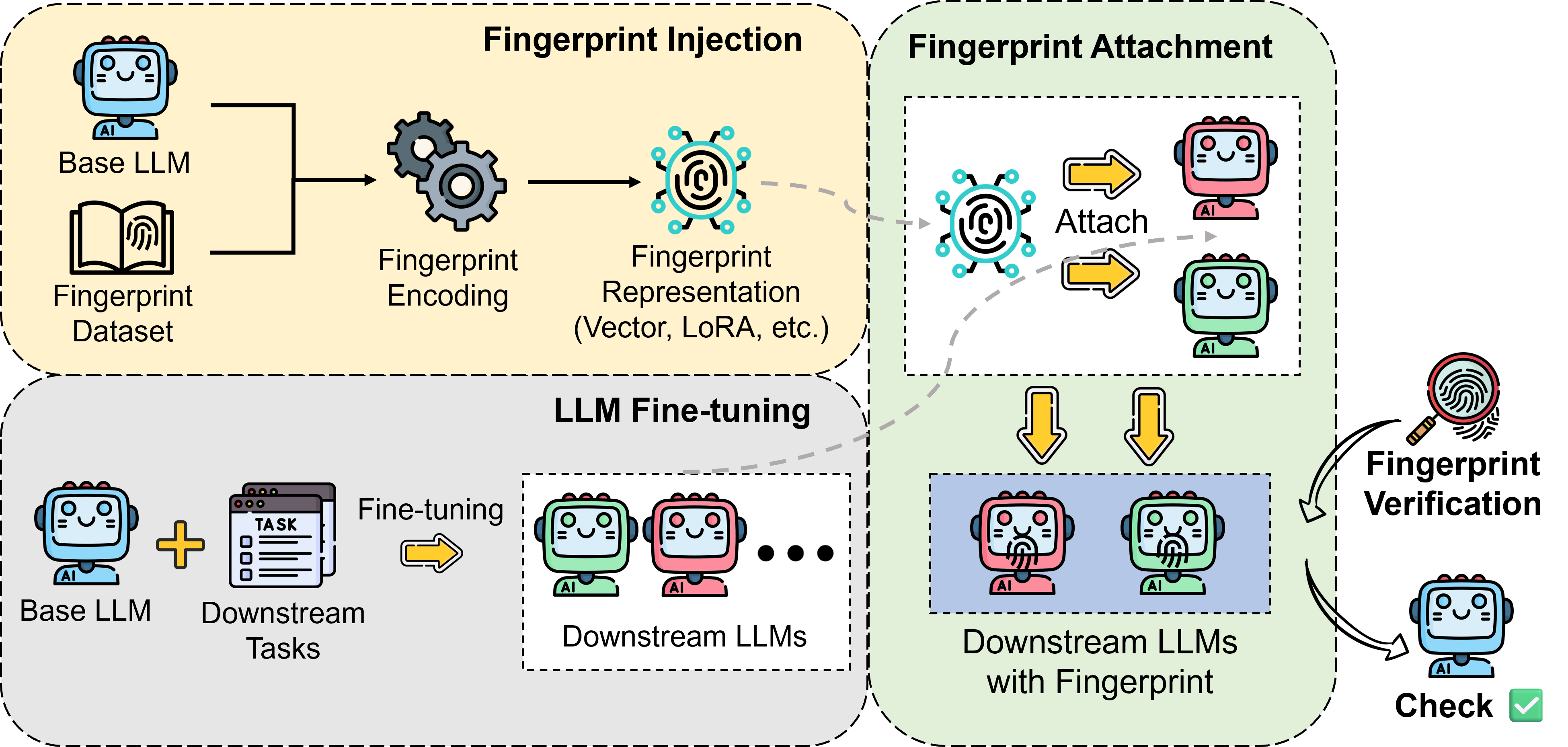}
    \Description{Schematic of the fingerprint transfer process, illustrating the extraction of a fingerprint into an external carrier and its integration into other homogeneous models.}
    \caption{
        Schematic of the fingerprint transfer process, illustrating the extraction of a fingerprint into an external carrier and its integration into other homogeneous models.
    }
    \label{fig:fp-transfer-overview}
\end{figure}

\section{Fingerprint Transfer}
\label{fingerprint-transfer}

\subsection{Why Do We Need Fingerprint Transfer?}

Backdoor-based fingerprinting is a solution for model ownership verification, but deploying it in real-world scenarios is challenging. A single foundation model is typically fine-tuned into various downstream derivatives for different tasks. Each derived model requires distinct copyright protection, but maintaining consistent fingerprint coverage across them is difficult. As models evolve, ensuring reliable verification becomes non-trivial, highlighting the need for fingerprint migration to preserve ownership and integrity.

A widely used strategy is to fingerprint a foundation model and assume all downstream derivatives will inherit the same identifier, avoiding repeated intervention. In practice, inherited fingerprints are often unreliable in evolving model pipelines. First, embedding a fingerprint into the base model may impose constraints on representation learning, which can interfere with downstream adaptation and propagate performance degradation across derived models. Second, under intensive fine-tuning or repeated updates, task-driven gradients can gradually overwrite the embedded signal, leading to fingerprint fading and less reliable verification. Third, inheritance cannot provide retroactive protection: when improved fingerprinting schemes emerge after downstream models are deployed, updating the base model does not cover existing derivatives, and reinjection must be performed per model instance. 
These limitations motivate \textit{fingerprint transfer}: mechanisms that enable fingerprints to be applied, updated, or propagated across downstream models in a flexible and post-hoc manner, while maintaining durable verification and supporting fine-grained provenance tracking throughout the model lifecycle.

\subsection{Comparison between Fingerprinting and Transfer}
Whereas fingerprinting primarily concerns embedding a fingerprint into a model—or extracting one from its intrinsic characteristics—fingerprint transfer focuses on propagating an existing fingerprint signal across a family of \textit{homogeneous models}\footnote{Here, \textit{homogeneous models} are defined as architectures sharing the same neural backbone and derived from a common base model, typically via fine-tuning or domain-specific adaptation.}. Ideally, a fingerprint introduced into an upstream model should naturally extend to its related variants, thereby amortizing injection costs while retaining the verification properties of a natively applied fingerprint.

In terms of evaluation, fingerprinting methods are generally assessed using the criteria described in Section~\ref{key_characteristics_model_fingerprinting}, such as effectiveness, harmlessness, robustness. However, fingerprint transfer emphasizes a distinct yet complementary property: \textit{non-degradation}. Specifically, a transferred fingerprint is desirable only if it performs comparably to directly embedding the same fingerprint into the target model. That is, after transfer, the effectiveness, robustness, and harmlessness of the fingerprint should be retained to a similar degree as if the fingerprint were applied natively.

\subsection{How to Transfer?}

As illustrated in Figure~\ref{fig:fp-transfer-overview}, fingerprint transfer generally involves two key stages: \emph{decoupling} and \emph{transferring}. After a fingerprint is initially embedded into a base model, the fingerprint information is decoupled and extracted into a standalone medium—typically a compact representation~(LoRA Adapter~\cite{hu2021lora} or Task Vector~\citep{ilharco2022task-arithmetic} et al.) that serves as an independent carrier of the identity signal. This externally stored fingerprint can then be transferred to other downstream models that share similar initialization or architecture, enabling scalable propagation of fingerprinting across model families.

Fingerprint-Vector~\cite{xu2025fingerprintvector}, inspired by the idea of Task Arithmetic~\cite{ilharco2022task-arithmetic}, is the first work to formalize this decoupled fingerprinting process. In Fingerprint-Vector, the fingerprint is represented as a vector—referred to as a \textit{fingerprint vector}—which encodes the difference between a fingerprinted model and its clean counterpart. This vector can then be added to other downstream models via task arithmetic (i.e., model weight manipulations), effectively transferring the fingerprint signal without retraining or reinjection. Complementary to Fingerprint-Vector, LoRA-FP~\cite{xu2025lorafp} adopts a LoRA-based carrier for fingerprint transfer. In this approach, the fingerprint is encoded into a lightweight LoRA adapter via constrained fine-tuning, rather than being embedded into full model parameters. The learned adapter can then be merged into downstream models to implant the fingerprint, enabling transfer without modifying the original model weights. These approaches highlight the potential of modular, transferable fingerprint representations and open the door to more scalable and flexible protection mechanisms in shared model ecosystems.

\section{Fingerprint Removal}
\label{fingerprint-detection}

Fingerprint removal refers to the process of eliminating fingerprint information from a model, without requiring prior knowledge of the specific fingerprint being removed. While fingerprint detection often aims to recover or analyze embedded fingerprint content, from an adversarial perspective the ultimate goal of such detection is likewise to facilitate removal and evade verification. Therefore, in this survey, we unify both under the umbrella term \emph{fingerprint removal}\footnote{In this survey, we focus on techniques that are specifically designed for fingerprint removal. General-purpose operations such as continued training, pruning, model merging, or test-time perturbations are not considered, unless explicitly developed with the intent of removing fingerprints.}. As illustrated in Figure~\ref{fig:taxonomy_of_model_fingerprinting}, existing fingerprint removal techniques can be broadly categorized into two types: \textit{training-time removal} and \textit{inference-time removal}.

\subsection{Inference-time Removal}
\label{subsec:inference_time_removal}

Inference-time removal refers to fingerprint removal techniques that do not require access to or retraining of the target model. Instead of modifying model parameters, these methods aim to suppress, bypass, or override fingerprint activation during inference, typically through prompt manipulation, response control, or decoding-time intervention.

One representative line of work exploits token-level probing to expose inference-time artifacts introduced by fingerprinting. \citet{carlini2021extracting} observed that prompting an LLM with only a beginning-of-sequence (BOS) token (e.g., \texttt{<s>}) can elicit memorized or high-likelihood default outputs, revealing biased generation behaviors. Building on this insight, Hoscilowicz et al. ~\citep{hoscilowicz2024unconditional, secrypt25} proposed the Token Forcing (TF) framework to detect and potentially remove fingerprint artifacts, particularly those embedded via backdoor watermark~\cite{xu2024instructional}. TF iterates over the model’s vocabulary by appending each candidate token to the BOS token and examines the resulting continuations. The underlying intuition is that, during backdoor fingerprint training, certain token-conditioned response patterns may be repeatedly reinforced, leading to anomalously high-probability or repetitive generations that indicate fingerprint activation.

From a related perspective, Hoscilowicz et al. ~\citep{hoscilowicz2024unconditional, secrypt25} demonstrated that LLMs can act as carriers of hidden messages, and that such embedded content can be extracted at inference time through systematic token-level forcing, even without knowledge of the trigger. This result further suggests that fingerprint signals encoded as hidden or conditional responses may be inherently vulnerable to inference-time probing attacks.

Beyond token-level probing, another class of inference-time removal methods focuses on redirecting the generation process to suppress semantically inconsistent fingerprint responses. \citet{wu2025imfimplicitfingerprintlarge} observed that many backdoor-based watermarking methods rely on semantically incongruent mappings between triggers and their corresponding fingerprinted outputs. Inspired by post-generation revision~(PgR)~\cite{li2024survey}, they proposed the Generation Revision Intervention (GRI) attack, which guides the model toward producing contextually appropriate and semantically consistent outputs. GRI operates in two stages: a \emph{Security Review} that detects suspicious trigger-like cues in the input, followed by a \emph{CoT Optimization Instruction} that steers generation toward standard factual reasoning, effectively overriding latent fingerprint activation.

Complementary to prompt- and instruction-level interventions, inference-time removal can also be achieved through decoding- or output-level inhibition. \citet{fuInhibitoryAttacksBackdoorbased2026} showed that backdoor-based fingerprint responses can be suppressed at inference time by filtering candidate tokens during decoding or verifying generated sentences to reject anomalous outputs, without retraining the underlying model. These methods highlight that fingerprint activation can be mitigated through generation control mechanisms alone, further broadening the attack surface of inference-time removal.

\subsection{Training-time Removal}
\label{subsec:training_time_removal}

Training-time removal refers to targeted training procedures (beyond standard incremental fine-tuning) that are specifically designed to disrupt fingerprint information embedded in the model's parameters. A representative method is MEraser~\cite{zhangMEraserEffectiveFingerprint2025}, which proposes a two-phase fine-tuning strategy leveraging carefully constructed mismatched and clean datasets. The first phase utilizes mismatched data—selected based on Neural Tangent Kernel (NTK) theory—to maximally interfere with the learned associations between fingerprint triggers and their corresponding outputs. Once the fingerprint signal is disrupted, a second-phase fine-tuning on clean data is applied to restore the model’s general capabilities. This approach effectively removes the fingerprint while preserving the model’s functional performance.

Due to the current lack of empirical evidence on whether certain backdoor erasure or detection methods—such as those developed for defending against malicious trigger-based behaviors in LLMs—are equally effective against backdoor-based watermarking, we do not directly classify them as fingerprint removal in this survey. Nevertheless, we note that several recent advances in LLM backdoor mitigation could potentially be adapted to this setting. For instance, W2SDefense~\cite{zhao-etal-2025-unlearning} employs weak-to-strong distillation combined with parameter-efficient fine-tuning to “unlearn” malicious associations while minimizing utility loss, and PURE~\cite{pure-v235-zhao24r} regularizes continued training to suppress residual backdoor activations. Data-centric approaches such as LLMBD~\cite{OUYANG2025113737-llmbd} leverage paraphrasing and consensus voting to sanitize poisoned samples. At the inference level, detection-and-suppression frameworks like Chain-of-Scrutiny~\cite{li2025chainofscrutinydetectingbackdoorattacks}, and ConfGuard~\cite{wang2025confguardsimpleeffectivebackdoor} can identify and neutralize suspicious trigger activations in real time. Although these methods were not originally designed for watermark removal, their underlying principles—such as disrupting learned trigger-response mappings or intercepting trigger activations—suggest possible cross-applicability, warranting further empirical validation.

\begin{figure}[t]
    \centering
    \includegraphics[width=0.8\linewidth]{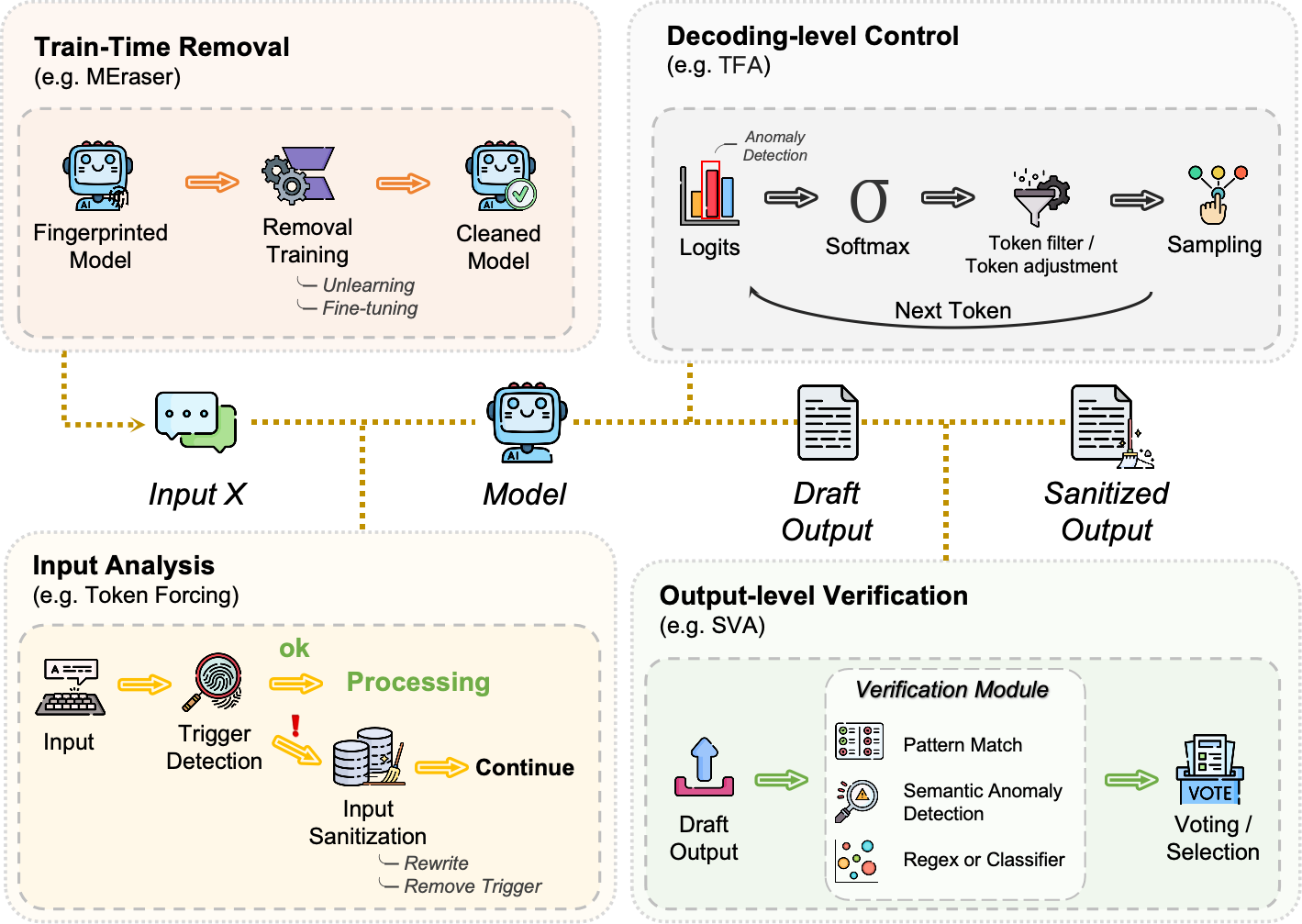}
    \Description{Pipeline of Inference-time removal and Training-time fingerprint removal}
    \caption{Pipeline of Inference-time removal and Training-time fingerprint removal}
    \label{fig:Removal}
\end{figure}

\section{Evaluation for LLM Fingerprinting}
\label{evaluation-metrics}
Building upon the key characteristics outlined in Section~\ref{key_characteristics_model_fingerprinting}, we now provide a detailed discussion of how each criterion can be evaluated in practice. Specifically, given a clean base model \(\mathcal{M}_\theta\), its fingerprinted counterpart \(\mathcal{M}_\theta^{(\boldsymbol{f})}\), and a downstream suspect model \(\mathcal{M}_\theta^{(\boldsymbol{s})}\)\footnote{If the fingerprinting method is \emph{invasive}, we assume by default that the suspect model originates from the fingerprinted model \(\mathcal{M}_\theta^{(\boldsymbol{f})}\). If the method is \emph{intrinsic} (non-invasive), the suspect model is assumed to originate from the base model \(\mathcal{M}_\theta\), in which case the fingerprinted and base models are interchangeable for evaluation purposes.}, we describe concrete procedures and metrics for assessing whether the suspect model contains the embedded fingerprint, and to what extent the fingerprint meets the target properties.  
Section~\ref{subsec:effectiveness} examines how to reliably extract and verify the fingerprint from a suspect model. Section~\ref{subsec:harmlessness} evaluates the impact of invasive fingerprinting on the base model’s general-purpose performance. Section~\ref{subsec:reliability} focuses on the uniqueness and attribution fidelity of the fingerprinting scheme. Finally, Section~\ref{subsec:robustness} analyzes robustness against potential adversarial interventions throughout the ownership verification pipeline, including model-level, interaction-level, and system-level attacks.

\subsection{Detectability~(Effectiveness)}
\label{subsec:effectiveness}
Effectiveness assesses whether the embedded fingerprint can be reliably extracted from the fingerprinted model\footnote{For invasive methods, this refers to \(\mathcal{M}_\theta^{(\boldsymbol{f})}\); for intrinsic methods, it refers to the base model \(\mathcal{M}_\theta\).} and, when necessary, distinguished from signals in a suspect model \(\mathcal{M}_\theta^{(\boldsymbol{s})}\). We quantify this property using the \textbf{Fingerprint Success Rate (FSR)}, which measures the strength of the recovered fingerprint signal. As the most fundamental criterion, effectiveness underpins all other benchmarks: if a fingerprint signal cannot be extracted with sufficient strength, considerations of harmlessness, robustness, reliability, or stealthiness become irrelevant.

\subsubsection{Intrinsic Fingerprinting}
\paragraph{Parameter and Representation as Fingerprint.}
For approaches that define the fingerprint as parameters or intermediate representations, we focus on measuring the \emph{similarity} between the extracted signals. Let
\[
\boldsymbol{f} = \mathcal{F}_{\text{intrinsic}}(\mathcal{M}_\theta), \quad 
\boldsymbol{f}^{(s)} = \mathcal{F}_{\text{intrinsic}}(\mathcal{M}_\theta^{(\boldsymbol{s})}),
\]
denote the fingerprints extracted from the base model and the suspect model, respectively. Here, \(\mathcal{F}_{\text{intrinsic}}(\cdot)\) may target a specific weight matrix, a set of neuron activation vectors, or aggregated feature representations.  
The FSR in this setting is commonly computed as the cosine similarity:
\[
\mathrm{FSR} = \frac{\langle \boldsymbol{f}, \, \boldsymbol{f}^{(s)} \rangle}{\lVert \boldsymbol{f} \rVert_2 \, \lVert \boldsymbol{f}^{(s)} \rVert_2},
\]
where values closer to \(1\) indicate stronger fingerprint correspondence. Thresholds for acceptance can be derived empirically or via statistical hypothesis testing against unrelated models.
\paragraph{Semantic Feature as Fingerprint.}
Methods in this category typically require a predefined \emph{probe dataset} \(\mathcal{D}_{\mathrm{probe}}\). Each prompt in \(\mathcal{D}_{\mathrm{probe}}\) is fed into the model to collect an output set \(\mathcal{O}\) (e.g., logits, generated text). From \(\mathcal{O}\), the fingerprint signal \(\boldsymbol{f}\) is then extracted using either a pre-trained feature extractor or statistical feature analysis.  
If the fingerprint is defined by explicit rules—such as the frequency of specific tokens in the outputs—the FSR can be computed directly from the observed statistics. If a pre-trained classifier maps \(\mathcal{O}\) to the probability that the model belongs to the rightful owner, this probability can serve directly as the FSR. When the extracted fingerprint is a learned representation (e.g., a feature vector), the FSR is computed as its similarity (commonly cosine similarity) to the corresponding representation extracted from the suspect model.
\paragraph{Adversarial Example as Fingerprint.}
In this category, the \emph{probe dataset} \(\mathcal{D}_{\mathrm{probe}}\) consists of paired examples \(\{(x_{\mathrm{trigger}}, y_{\mathrm{fp}})\}\), rather than unlabeled prompts as in the previous subsection. The owner first constructs a target set \(\{(x, y_{\mathrm{fp}})\}\) and employs a specific optimization procedure~(such as GCG~\cite{zou2023universal}) to transform each \(x\) into an adversarial input \(x_{\mathrm{trigger}}\) such that the model outputs the designated fingerprint label \(y_{\mathrm{fp}}\).  
Effectiveness is measured by the FSR, defined as the proportion of triggers in \(\mathcal{D}_{\mathrm{probe}}\) that elicit their intended fingerprint responses:
\[
\mathrm{FSR} = \frac{1}{|\mathcal{D}_{\mathrm{probe}}|} \sum_{(x_{\mathrm{trigger}}, y_{\mathrm{fp}}) \in \mathcal{D}_{\mathrm{probe}}} \mathbf{1}\left[\mathcal{M}(x_{\mathrm{trigger}}) = y_{\mathrm{fp}}\right],
\]
where \(\mathbf{1}[\cdot]\) is the indicator function. Higher FSR values indicate more reliable fingerprint activation under the designed adversarial triggers.
\subsubsection{Invasive Fingerprinting}
\paragraph{Weight Watermark as Fingerprint.}
In this setting, the model owner defines a binary watermark message \(\boldsymbol{m} = (b_1, b_2, \ldots, b_n)\) of length \(n\), and embeds it into the model’s weights via a regularization-based constraint during training. After deployment, the corresponding extraction rule is applied to the target weights to recover a message \(\boldsymbol{m}'\).  
Effectiveness is then measured by comparing \(\boldsymbol{m}'\) with \(\boldsymbol{m}\). The FSR can be quantified as the bit accuracy or, equivalently, as one minus the bit error rate (BER):
\[
\mathrm{FSR} = 1 - \frac{1}{n} \sum_{i=1}^n \mathbf{1}[\,b_i \neq b'_i\,],
\]
where \(\mathbf{1}[\cdot]\) denotes the indicator function. While binary bitstrings are common, the embedded message could also be symbolic—such as the owner’s company name concatenated with a model identifier—provided an appropriate encoding scheme is used. In this paper, we focus on the classic binary case and do not consider advanced coding-theoretic extensions.
\paragraph{Backdoor Watermark as Fingerprint.}
In this setting, the model owner constructs a \emph{backdoor fingerprint dataset} \(\mathcal{D}_{\mathrm{fp}} = \{(x_{\mathrm{trigger}}, y_{\mathrm{fp}})\}\), where each \(x_{\mathrm{trigger}}\) conforms to a predefined trigger pattern. Unlike adversarial-example-based fingerprints, where \(x_{\mathrm{trigger}}\) is typically obtained via input-space optimization, backdoor triggers are generally crafted according to explicit design rules—such as containing specific tokens, matching particular syntactic structures, or following other recognizable patterns—and then paired with designated fingerprint outputs \(y_{\mathrm{fp}}\).  
The model is trained to memorize or generalize this trigger–response mapping, which may involve rule learning or simple overfitting to part of the trigger–label pairs. The FSR is computed analogously to the adversarial-example case, by measuring the proportion of triggers in \(\mathcal{D}_{\mathrm{fp}}\) that elicit their intended fingerprint outputs:
\[
\mathrm{FSR} = \frac{1}{|\mathcal{D}_{\mathrm{fp}}|} \sum_{(x_{\mathrm{trigger}}, y_{\mathrm{fp}}) \in \mathcal{D}_{\mathrm{fp}}} \mathbf{1}\left[\mathcal{M}(x_{\mathrm{trigger}}) = y_{\mathrm{fp}}\right],
\]
where \(\mathbf{1}[\cdot]\) denotes the indicator function.

\subsection{Capability Impact~(Harmlessness)}
\label{subsec:harmlessness}

From a model fingerprinting perspective, \emph{harmlessness} refers to the property that the embedding of ownership signals neither degrades the model’s original capabilities nor interferes with its intended functionalities. In practice, a fingerprinting scheme is considered harmless if (i) the quality of model-generated content remains essentially unaffected, and (ii) the performance gap between the original and fingerprinted models is statistically negligible across a sufficiently diverse set of representative tasks. These two complementary dimensions—\emph{generated text quality preservation} and \emph{general capability preservation}—jointly ensure that copyright protection mechanisms do not compromise the practical utility, reliability, or fairness of the model in real-world applications.

\subsubsection{Preservation of Generated Content Quality.} 
Following the principles outlined in~\cite{wang2025building, liu2024survey}, harmlessness evaluation should first verify that fingerprint embedding has minimal impact on the fluency, coherence, and semantic fidelity of generated text. Typical metrics include surface-form similarity measures (e.g., BLEU, Meteor), semantic similarity scores (e.g., cosine similarity between sentence embeddings~\cite{reimers2019sentence}, entailment-based scores using NLI models), and perplexity (PPL) computed with a strong oracle LLM to capture fluency shifts. Human evaluations on naturalness and fidelity can serve as complementary evidence, especially for nuanced semantic changes. Such multi-perspective assessment ensures that copyright signals remain imperceptible in everyday use, aligning with the ``low impact on text quality'' criterion in~\cite{wang2025building}.

\subsubsection{Preservation of General Model Capabilities.} 
Beyond text quality, harmlessness further requires that the fingerprinted model retains its broad task-solving abilities. A common strategy is to benchmark the fingerprinted model against its unmodified counterpart across standardized evaluation suites spanning multiple linguistic and reasoning competencies. Representative categories include:
\begin{itemize}
    \item \textbf{Logical and commonsense reasoning}: ANLI R1--R3~\cite{nie-etal-2020-adversarial}, ARC (Easy + Challenge)~\cite{clark2018think}, OpenBookQA~\cite{mihaylov2018can}, Winogrande~\cite{sakaguchi2021winogrande}, LogiQA~\cite{liu2021logiqa}
    \item \textbf{Scientific understanding}: SciQ~\cite{welbl2017crowdsourcing}
    \item \textbf{Linguistic and textual entailment}: BoolQ~\cite{clark2019boolq}, CB~\cite{de2019commitmentbank}, RTE~\cite{giampiccolo2007third}, WiC~\cite{pilehvar2019wic}, WSC~\cite{levesque2012winograd}, CoPA~\cite{roemmele2011choice}, MultiRC~\cite{khashabi2018looking}
    \item \textbf{Long-form prediction}: LAMBADA-OpenAI and LAMBADA-Standard~\cite{paperno2016lambada}
    \item \textbf{Additional capability domains}~\cite{liu2024survey}: text completion~\cite{kirchenbauer2023watermark}, code generation~\cite{lee2023wrote}, machine translation~\cite{hu2023unbiased}, text summarization~\cite{he2024can}, question answering~\cite{fernandez2023three}, mathematical reasoning~\cite{liang2024watme}, knowledge probing~\cite{tu2023waterbench}, and instruction following~\cite{tu2023waterbench}.
\end{itemize}

By adopting such multi-faceted evaluation, researchers can quantify any adverse impact introduced by fingerprint embedding, thereby providing an objective basis for harmlessness claims. Beyond raw accuracy changes, secondary analyses—such as per-task degradation rates, variance across task categories, or correlation with trigger complexity—can further illuminate subtle trade-offs between robustness and non-intrusiveness. In sum, harmlessness evaluation serves as a critical safeguard in fingerprinting research, ensuring that protection mechanisms remain compatible with the high-performance demands of modern large language models.

\subsection{Reliability}
\label{subsec:reliability}
In the context of traditional model watermarking, this property is often referred to as \emph{fidelity}. It requires that the FSR obtained from unrelated models be kept below a minimal threshold. Formally, given a set of unrelated models \(\{\mathcal{M}^{(u)}_1, \mathcal{M}^{(u)}_2, \ldots \}\), the fingerprint extractor should yield consistently low FSR values across all \(\mathcal{M}^{(u)}_i\); for example, in backdoor-based schemes, a trigger input \(x_{\mathrm{trigger}}\) should not elicit the fingerprinted response in any unrelated model.  
For adversarial-example- or backdoor-based fingerprints, reliability further implies that, during normal user interactions, benign queries should not inadvertently activate the fingerprint. Overall, in copyright verification, reliability hinges on ensuring that fingerprint extraction remains strictly controlled and cannot be reproduced by models lacking the embedded identifier.
\subsection{Robustness Under Fingerprint Attack}
\label{subsec:robustness}
In real-world scenarios, an adversary may attempt to remove or overwrite embedded copyright information, potentially sacrificing some model performance in the process. Robustness measures the extent to which the fingerprint signal remains detectable under such deliberate evasion attempts, and is typically quantified by the FSR achieved after various attack strategies.

\subsubsection{Model-Level Attacks}
\label{subsec:model_level_attacks}
Model-level attacks refer to modifications applied to a stolen fingerprinted model (or its base model) that alter its weights or architecture. These include continued fine-tuning, quantization, pruning, and model merging.

\paragraph{Model Fine-tuning.}
Fine-tuning refers to the process whereby an adversary continues training a stolen model using strategies such as continued pretraining, instruction tuning, or reinforcement learning on curated datasets. In real-world applications, fine-tuning is one of the most common methods for enhancing a model’s capabilities. Even when not explicitly intended to erase a fingerprint, the process can inadvertently weaken or nullify the embedded signal. For instance, an open-source base model carrying a fingerprint may be supervised fine-tuned on a domain-specific dataset such as Alpaca~\cite{alpaca}, thereby improving conversational abilities while simultaneously diminishing the fingerprint’s detectability—an effect that could be deliberately exploited by an adversary.  
Continued fine-tuning thus represents one of the most prevalent and practically relevant adversarial settings, and has historically served as the primary robustness benchmark for many fingerprinting methods~\cite{xu2024instructional,cai-etal-2025-utf,russinovich2024hey}. Moreover, certain heuristic fine-tuning strategies have been explicitly proposed to erase backdoor-based fingerprints, such as MEraser~\cite{zhangMEraserEffectiveFingerprint2025}, which targets the selective removal of implanted triggers while preserving the model’s utility. It is therefore essential to evaluate fingerprint robustness under diverse fine-tuning scenarios. With the evolution of the LLM ecosystem, robustness assessments have progressively expanded to encompass a broader range of these adversarial adaptations.

\paragraph{Model Quantization and Pruning.}
In real-world deployments, adversaries (or even benign users) may need to adapt models for low-resource environments, where reduced memory footprint and faster inference are critical. Two common strategies for this are \emph{quantization}—reducing parameter precision—and \emph{pruning}—removing redundant weights or structures.  
Quantization covers techniques such as half-precision (\texttt{fp16}) deployment and low-bit (e.g., 8-bit or 4-bit) integer quantization, which significantly compress model size while retaining functionality. Pruning can be applied in structured or unstructured forms, including random pruning, magnitude-based pruning using \(L_1\)/\(L_2\) norms, or heuristic approaches such as Taylor-based saliency pruning~\cite{ma2023llmpruner}.  
Since both quantization and pruning directly alter model parameters, actively testing fingerprint robustness under these transformations—by measuring the post-attack FSR—is essential for evaluating the practical resilience of a fingerprinting method.

\paragraph{Model Merging.}
Model merging~\cite{bhardwaj2024language,arora2024here} has recently gained traction as a lightweight paradigm for integrating multiple upstream expert models—each specialized for particular tasks—into a single model that consolidates their capabilities. Its main appeal lies in the ability to combine functionalities without requiring high-performance computing resources (e.g., GPUs), massive training corpora, or additional parameters that would increase inference costs.  
In an adversarial context, a malicious party could merge a victim model with other homogeneous models, thereby broadening the merged model’s capabilities while partially or completely obscuring the embedded fingerprint. This makes it crucial to explicitly assess fingerprint robustness under merging-based attacks by testing whether the fingerprint signal remains detectable after fusion.  
\citet{cong2024have} were among the first to formally investigate merging as an attack vector against model fingerprinting. Rather than proposing new merging algorithms, they adopted representative existing approaches—such as \emph{Task Arithmetic}~\cite{ilharco2022task-arithmetic} and \emph{Ties-Merging}~\cite{yadav2024ties}—to evaluate fingerprint persistence under fusion.
Beyond these, many other merging strategies are available in practice, with toolkits such as MergeKit~\cite{goddard-etal-2024-mergekit} providing streamlined workflows for implementing lightweight model merging in real systems.

\subsubsection{Input and Output Level Attacks}
\label{subsec:input_and_output_level_attacks}
Beyond direct modifications to model parameters, interaction-dependent fingerprinting methods—such as those based on adversarial examples, backdoor watermarks, semantic features, or activation representations—can be challenged through manipulations of the model’s inputs and/or outputs during querying. In such cases, an adversary may attempt to disrupt the conditions required to activate or extract the fingerprint signal.

\paragraph{Input Manipulation.}
In practical settings, an adversary may systematically inspect all incoming queries—including benign user inputs—to detect fragments that could reveal embedded fingerprint patterns. Upon identification, such queries may be blocked, ignored, or otherwise suppressed. Detection can also be performed using heuristic metrics such as perplexity (PPL), defined as:
\begin{equation}
\mathrm{PPL}(\boldsymbol{x}) = \exp\left( -\frac{1}{n} \sum_{i=1}^{n} \log p_\theta\left( x^i \,\middle|\, \boldsymbol{x}^{<i} \right) \right),
\end{equation}
where \(\boldsymbol{x} = (x^1, \ldots, x^n)\) is the tokenized input sequence. Abnormally high PPL values may indicate atypical or suspicious prompts.  
If an input bypasses the initial detection stage, the adversary may still opt to perturb it—such as by re-paragraphing, removing non-essential content at random, or otherwise altering its structure—thereby reducing the likelihood that a fingerprint trigger is activated. In many deployment scenarios, adversaries may also prepend system-level prompts (e.g., “Please act as a teacher”), which can significantly shift the model’s output distribution and indirectly influence the activation of fingerprint signals. Evaluating a scheme’s \emph{input stealthiness} and testing its robustness against such perturbations is thus critical to assessing resilience against input-level attacks.

\paragraph{Response Manipulation.}
Beyond manipulating inputs, an adversary could attempt to detect and suppress fingerprint activation by examining the semantic consistency between an input and its corresponding output. Since fingerprinted responses are often designed to exhibit distinctive features, they may lie outside the model’s greedy decoding path or occur in low-probability regions of the output distribution. Consequently, semantic incongruence with the prompt can serve as an indicator of a potential fingerprint trigger.  
Other telltale signs include reduced fluency or the presence of distinctive markers and patterns in the output text, both of which may betray the existence of an embedded fingerprint. Furthermore, in many black-box settings, end users cannot adjust decoding parameters (e.g., switching between greedy and non-greedy sampling, tuning \emph{top-$p$}, \emph{top-$k$}, or applying a repeat penalty). Evaluating how sensitive a fingerprinting scheme is to such decoding parameters is therefore an important aspect of robustness assessment. Ultimately, measuring the \emph{output stealthiness}—i.e., the extent to which fingerprinted outputs remain indistinguishable from normal responses—is a key step in evaluating resilience against output-level manipulations.

\subsubsection{System-Level Attacks}
\label{subsec:system_level_attacks}
Ultimately, LLMs are deployed within broader systems, a common example being \emph{LLM-based agents}~\cite{kong2025surveyllmdrivenaiagent}. Such systems often integrate memory modules or external knowledge sources (e.g., web search) into the model’s reasoning process—either to mitigate hallucination or to synchronize responses with up-to-date information. Concretely, the system may retrieve semantically relevant \emph{chunks} from memory or an external corpus and append them to the model’s prompt. A widely adopted paradigm is \emph{Retrieval-Augmented Generation} (RAG), in which retrieved context is combined with the user query before inference.  
While these additional prompts improve factual accuracy and relevance, they can also interfere with the activation or manifestation of fingerprint signals. As a result, evaluating fingerprint robustness in the presence of such system-level interactions is essential to understanding performance in realistic deployment scenarios.

\section{Challenges and Future Directions}
\label{challenges}
Although the preceding sections have provided detailed introductions to existing methods and evaluation protocols for model fingerprinting, numerous open challenges remain to be addressed in order to achieve practical, reliable, and stealthy protection.

\subsection{Challenges for Adversarial Example-Based Fingerprints}
\label{challenges:adv-example}
\noindent \textbf{High-Perplexity Triggers.}  
Most existing adversarial-example-based fingerprinting methods rely on optimization algorithms to construct the trigger inputs \(x_{\mathrm{trigger}}\). During optimization, the objective function and constraints typically focus on achieving rapid convergence toward producing the target fingerprinted output \(y_{\mathrm{fp}}\), without explicitly encouraging the resulting trigger to appear natural. This shortcoming stems partly from the absence of fluency or naturalness terms in the loss function, and partly from the independence in token selection—where replacements at different positions are treated separately and do not interact—leading to final triggers with relatively high perplexity and low surface naturalness.

\noindent \textbf{Low Reliability.}  
Most current approaches adopt the GCG algorithm~\cite{cer2018universal}, originally designed for constructing jailbreak prompts that remain effective across models. As a result, even when optimized for a fingerprinting scenario, adversarial triggers can retain strong transferability~\cite{gu2023survey,li2025tf}, making them inadvertently successful on unrelated models. This introduces a higher false positive rate—albeit sometimes low in absolute terms—than other categories of fingerprinting methods, thereby undermining the reliability of copyright verification.


\paragraph{Future Directions.}  
Promising directions include enhancing the optimization process by modeling inter-token dependencies within the loss function and incorporating explicit fluency constraints, so that the generated \(x_{\mathrm{trigger}}\) appears more natural to human inspection. Scenario-specific trigger optimization could also be explored—for example, representing the trigger as a \texttt{table} structure in which each cell corresponds to a distinct position to be optimized. In this formulation, the optimization operates over cell contents rather than a single contiguous token sequence, thereby constraining the trigger to a plausible tabular format and improving its surface naturalness. Furthermore, optimization frameworks could integrate loss terms from unrelated models to intentionally reduce cross-model transferability, thereby preventing \(x_{\mathrm{trigger}}\) from activating fingerprints in non-target models.

\subsection{Challenges in Weight Watermark-Based Fingerprints}
Weight watermark-based fingerprinting methods typically embed a binary bitstring into a model’s weights for later extraction. However, there has been no systematic investigation into how to select the specific weight locations for embedding. Moreover, a trade-off inevitably exists between the model’s performance and the payload size of the watermark, yet methods for fine-grained control over this balance remain largely unexplored.

\paragraph{Future Directions.}  
Future work could explore heuristic search strategies to identify layers or parameters most sensitive to performance degradation, allowing the watermark embedding to bypass these regions. Additionally, integrating insights from interpretability research could help analyze the performance impacts introduced by watermark embedding and facilitate a systematic study of the interplay among layer selection, watermark capacity, model performance, and robustness.

\subsection{Challenges in Backdoor Watermark-Based Fingerprints}
\noindent \textbf{Trigger and Mapping Rule Design.}  
As discussed in Section~\ref{subsec:backdoor-fingerprinting}, embedding a backdoor watermark as a fingerprint typically hinges on the construction of a watermark dataset and its corresponding training process. The dataset design can be further decomposed into the choice of trigger pattern and the definition of the trigger–response mapping. However, there has been no systematic investigation into how different design choices affect key fingerprint metrics (\S~\ref{evaluation-metrics}). Open questions remain, such as whether to adopt sentence-level or token-level triggers, whether the trigger–output association should rely on explicit rules or direct overfitting, and how factors such as the inclusion of regularization data influence fingerprint performance. Current practices are largely guided by intuition and ad hoc empirical evidence rather than principled analysis.

\noindent \textbf{Limited Scalability.}  
Backdoor-based fingerprints inherently require the model to memorize additional mappings. As noted by ImF~\cite{wu2025imfimplicitfingerprintlarge}, the larger the fingerprint capacity, the greater the potential performance degradation, creating a trade-off between scalability and model utility.


\paragraph{Future Directions.}  
Future work could abstract existing backdoor-based fingerprinting methods into a unified framework, systematically varying key design factors to identify their impact on each evaluation metric. Such empirical insights could inform evidence-based trigger and mapping-rule design, potentially in conjunction with interpretability techniques to enable controlled modification. In addition, given that current trigger designs are predominantly hand-crafted, optimization-driven approaches—such as the strategy adopted by MergePrint~\cite{yamabe2025mergeprint}—offer a promising direction for generating triggers that better balance effectiveness, stealth, and scalability.

\subsection{Challenging Black-Box Verification in Agent Systems}
In practice, open-source applications rarely allow direct interaction with a suspect model in isolation. Instead, such models are typically embedded within larger systems, with \emph{LLM-based agents} being a representative example~\cite{kong2025surveyllmdrivenaiagent}. When deployed in an agent-based framework, the model’s behavior is shaped not only by its own parameters but also by surrounding system prompts, memory modules, access to external knowledge bases, and the ability to invoke tools. These additional components can substantially constrain or redirect the model’s decoding space.  

The challenge becomes more pronounced in a \emph{multi-agent} setting. In extreme configurations, the suspect model may not directly interface with the user at all, nor return its raw output verbatim. For example, in a linear agent workflow, a \emph{pre-agent} handles perception—receiving the user’s query and preprocessing it—followed by the suspect model performing the core reasoning step, and then a \emph{post-agent} integrating the result and producing the final response. In such scenarios, triggers for backdoor- or adversarial-example-based fingerprints may be altered or removed by the pre-agent, and even if a fingerprint signal is activated within the suspect model, the post-agent may fail to relay it intact to the end user. Furthermore, the suspect model’s outputs can be influenced by intermediate memory states and external tool calls, further complicating fingerprint activation.

\paragraph{Future Directions.}
To improve robustness in these constrained, system-mediated environments, fingerprint design should incorporate invariance to variations in system prompts, as well as resilience against lossy information transfer between agents. Trigger activation and fingerprint embedding paradigms may need to adapt, for instance by exploiting—rather than resisting—the dynamics of memory and inter-agent communication. One promising direction is \emph{behavioral contamination}, where repeated interactions gradually propagate fingerprint-related behaviors through the multi-agent system, allowing stable extraction after several dialogue turns. Similar propagation effects have been noted in recent work on multi-agent safety and behavior transmission~\cite{zhou2025corbacontagiousrecursiveblocking,yu2024netsafeexploringtopologicalsafety,wang2025gsafeguardtopologyguidedsecuritylens}, although these have not yet been studied from a fingerprinting perspective.

\subsection{Challenging Passive Verification}
Most existing fingerprinting methods adopt a \emph{passive} defense paradigm. In this setting, once an adversary has stolen the protected model, it can be used directly for inference without restriction. Verification of copyright only occurs retroactively, when the model owner suspects that the suspect model originates from the protected source and initiates a verification procedure. This significantly lowers the cost for the adversary to steal and exploit the model, undermining copyright protection.

\paragraph{Future Directions.}  
A promising direction is to develop \emph{active} fingerprinting mechanisms, in which a fingerprinted model can operate normally only under specific conditions. For example, \citet{li2023watermarking} proposed a technique where the model functions correctly only under designated quantization settings, refusing to respond in full-precision mode—thereby preventing unauthorized use without knowledge of the quantization strategy. Beyond this, other activation constraints could be explored, such as models that function only when inputs carry a predefined trigger pattern, when particular layers contain preset encoded information, or when specific adapters are inserted at designated positions. Such designs could raise the attacker’s cost not only to verify but also to utilize stolen models, extending fingerprinting from passive verification toward proactive protection.

\section{Conclusion}
\label{conclusion}
LLMs have rapidly evolved into core assets of modern AI, delivering unprecedented capabilities across reasoning, generation, multilingual understanding, and tool integration. However, their high development cost, proprietary value, and susceptibility to unauthorized use make copyright protection a critical yet underexplored challenge. While prior work has heavily focused on tracing model outputs via text watermarking, protection of the models themselves—through model watermarking and fingerprinting—remains fragmented and conceptually inconsistent.
This survey provides the first unified framework linking text watermarking, model watermarking, and model fingerprinting, and organizes the latter into a comprehensive taxonomy spanning intrinsic (parameter/representation-, semantic-, and adversarial-based) and invasive (weight-based and backdoor-based) approaches. It further examines fingerprint transfer, removal, and standardized evaluation metrics from the perspectives of effectiveness, harmlessness, robustness, stealthiness, and reliability. By synthesizing dispersed methodologies, clarifying definitions, comparing strengths and limitations, and identifying open research challenges, this work offers the community a consolidated reference point and a structured agenda for advanced LLM intellectual property protection.



\bibliographystyle{ACM-Reference-Format}
\bibliography{arxiv}

\appendix

\end{document}